\pdfoutput=1
\documentclass[a4paper,11pt]{article}

\newcommand\papertitle{Fermion Parity Resolution of Entanglement}

\usepackage{jheppub}
\usepackage[dvipsnames]{xcolor}

\usepackage{tikz}
\usetikzlibrary{arrows, backgrounds, calc, decorations.pathreplacing, decorations.markings, fadings, quotes, positioning, shadings, intersections, angles}

\usepackage{amsmath,amsfonts,amssymb,mathrsfs,dsfont}
\usepackage{mathtools}
\usepackage{bbm}
\usepackage[inline]{enumitem}
\usepackage{multirow}
\usepackage{subcaption}
\usepackage{nicefrac}
\usepackage{xspace}
\usepackage{MnSymbol}
\usepackage{comment}

\usepackage[colorlinks=true]{hyperref}
\hypersetup{
    bookmarks=true,         
    unicode=false,          
    pdftoolbar=true,        
    pdfmenubar=true,        
    pdffitwindow=false,     
    pdfstartview={FitH},    
    pdftitle={\papertitle}, 
    pdfauthor={Christian Northe},
    pdfnewwindow=true,      
    colorlinks=true,        
    linkcolor=teal,         
    citecolor=Maroon,          
    filecolor=magenta,      
    urlcolor=teal          
}

\makeatletter

\@addtoreset{equation}{section}
\makeatother

\newcommand{\p}	{\partial}
\newcommand{\bp}{\bar{\partial}}

\newcommand{\Z}	{\mathbb{Z}}
\newcommand{\N}	{\mathbb{N}}

\newcommand{\cC}{\mathcal{C}}

\newcommand{\cH}{\mathcal{H}}

\newcommand{\cO}{\mathcal{O}}

\newcommand{\cR}{\mathcal{R}}
\newcommand{\cS}{\mathcal{S}}
\newcommand{\cT}{\mathcal{T}}

\newcommand{\cZ}{\mathcal{Z}}

\DeclareMathOperator{\tr}{\tr}
\DeclareMathOperator{\Pf}{Pf}

\renewcommand{\tr}	{\mathrm{tr}}

\newcommand{\diag}  {\mathrm{diag}}

\newcommand{\bra}[1]	{\langle{#1}\vert}
\newcommand{\ket}[1]	{\vert{#1}\rangle}
\newcommand{\braket}[2]	{\langle{#1}\vert{#2}\rangle}
\newcommand{\corr}[1]   {\left\langle{#1}\right\rangle}

\newcommand{\bz}    {\bar{z}}

\newcommand{\bpsi}  {\bar{\psi}}

\newcommand{\bL}    {\bar{L}}
\newcommand{\bh}    {\bar{h}}

\newcommand{\bT}    {\bar{T}}
\newcommand{\bi}    {\bar{\iota}}

\newcommand{\tw}    {\tilde{w}}

\newcommand{\tq}    {\tilde{q}}
\newcommand{\ttau}  {\tilde{\tau}}

\newcommand{\ketbra}[2]	{\ket{#1}\bra{#2}}

\newcommand{\End}	{\mathrm{End}}
\newcommand\id		{\mathbf{1}}

\newcommand\Algebra[1]	{\mathfrak{#1}}

\newcommand{\su}	{\Algebra{su}}

\newcommand{\bbra}[1]	{\langle\!\langle{#1}\lVert}
\newcommand{\bket}[1]	{\lVert{#1}\rangle\!\rangle}

\newcommand{\gf}        {\mathsf{g}}
\newcommand{\rc}        {\mathsf{R}}

\newcommand{\Ot}        {\cO^{\textrm{top}}}
\newcommand{\Ob}        {\cO^{\textrm{bdy}}}

\newcommand{\modS}      {\cS}

\newcommand{\cc}        {\mathsf{c}}

\newcommand{\iu}{\mathsf{i}}

\newcommand{\ab}        { {\alpha\beta} }
\renewcommand{\aa}        { {\alpha\alpha} }

\newcommand{\wi}        {\mathsf{W}} 
\newcommand{\X}         {X}
\newcommand{\diffsn}[1] {\Delta s_{#1}}
\newcommand{\prob}      {\mathsf{p}}

\newcommand{\cft}{\mathrm{CFT}}

\renewcommand\emph[1]  {\textbf{#1}}

\newcommand\Secref[1]	{Section~\ref{#1}\xspace}

\newcommand\secref[1]	{section~\ref{#1}\xspace}

\newcommand\figref[1]	{figure~\ref{#1}\xspace}
\newcommand\appref[1] {appendix~\ref{#1}\xspace}

\begin{document}
\title{\papertitle}

\author{Christian Northe}

\affiliation{CEICO, Institute of Physics of the Czech Academy of Sciences,\\
Na Slovance 2, 182 00 Prague 8, Czech Republic}

\emailAdd{northe@fzu.cz}

\abstract{%
\begin{abstract} 

Entanglement is analyzed in the Majorana fermion conformal field theory (CFT) in the vacuum, in the fermion state, and in states built from conformal interfaces. In the boundary-state approach, the Hilbert space admits two factorizations for a single interval, producing distinct entanglement spectra determined by spin structures. Although Rényi and relative entropies are shown to be insensitive to these structures, symmetry-resolved entanglement naturally reveals their differences. The Majorana fermion's $\mathbb{Z}_2^F$ symmetry, generated by the fermion-parity operator $(-1)^F$, distinguishes bosonic from fermionic sectors, motivating the notion of fermion-parity resolution. While $\mathbb{Z}_2^F$ is naturally a symmetry of the vacuum and fermion reduced density matrices, the Hilbert space factorization is shown to stabilize this symmetry in conformal interface states. When an unpaired Majorana zero mode is present, fermion-parity-resolved entropies display equipartition at all orders in the UV cutoff; in its absence, the breaking of equipartition is quantified by Ramond-sector data. This behavior persists across all states considered. Connections with symmetry-protected topological phases of matter are outlined. All results are compared with twist field computations.
\end{abstract}

}

\maketitle
\flushbottom
\newpage
\section{Introduction}
\label{secIntro}
Majorana fermions occupy a central role in theoretical physics. As the simplest fermionic quantum field, they provide a foundational model for field-theoretic techniques \cite{DiFrancesco:1997nk} and appear across diverse contexts. They arise as worldsheet fields in superstring theory \cite{Recknagel:2013uja}, in models of neutrino mass generation and dark matter \cite{dutta2021singlet}, and as low-energy modes of the Kitaev chain \cite{kitaev2001unpaired}, a canonical symmetry-protected topological (SPT) phase. SPTs are $d$-dimensional gapped quantum field theories with a global symmetry $G$ that cannot be deformed to the trivial theory without breaking $G$ or closing the gap.

The Majorana fermion theory has recently been identified as the simplest fermionic minimal model \cite{hsieh2021fermionic, kulp2021two}, playing a role analogous to the Ising CFT for bosonic theories. Unlike bosonic CFTs, fermionic CFTs require not only a Riemann surface and metric but also a choice of spin structure. Fermionic minimal models have quickly proven important for SPTs \cite{smith2021boundary, boyle2022conformal}. The key observation is that transitions between trivial and non-trivial SPT phases with the same symmetry group $G$ trap unpaired Majorana zero modes, echoing the Jackiw–Rebbi mechanism \cite{jackiw1976solitons}. More generally, massive deformations of CFTs correspond to boundary states encoding boundary conditions \cite{han2017boundary, Cho:2016xjw}, and in fermionic minimal models these boundary states systematically trap Majorana zero modes \cite{smith2021boundary}.

Topological phases are distinguished by their entanglement structure \cite{ludwig2015topological}. In particular, their entanglement spectrum reflects the structure of physical edge modes \cite{PhysRevLett.108.196402,PhysRevLett.101.010504}. The purpose of this work is to study entanglement, and especially entanglement spectra, in fermionic minimal models through their simplest incarnation: the Majorana fermion theory. Being free, it affords exact treatment of fermionic characeristics and their impact on entanglement measures. Beyond primary states such as the vacuum $\ket{0}$ and fermion excitation $\ket{\psi}$, also more exotic excitations, namely states built from conformal interfaces \cite{bachas2012worldsheet, bachas2013fusion, Brehm:2015lja} are analyzed.

In passing it is noted that much work has been dedicated to entanglement in the Dirac fermion \cite{casini2005entanglement,casini2009entanglement, herzog2013entanglement, fries2019entanglement, Fries:2019ozf, bonsignori2019symmetry}, though unlike the Majorana fermion theory, it does not form a rational CFT, and thus not a fermionic minimal model. 

A secondary aim is to provide a case study of entanglement in excited states—both primary and beyond—within the boundary state approach to entanglement \cite{Ohmori_2015, cardy2016entanglement, roy2025boundary}, and to compare this with the standard twist field formalism\footnote{Both the uniformization map and twist field techniques are referred to as the “twist field formalism” in the following.} \cite{ibanez2012entanglement}. A partial analysis along these lines was initiated in \cite{yan2025symmetry} but not seen through. The framework of \cite{Ohmori_2015} was introduced to provide a meaningful factorization of the Hilbert space over spatial subregions in quantum field theory. Beyond this formal motivation, the boundary state approach affords direct control over the Hilbert spaces $\cH_A$ associated with subregions $A$. In the case of the Majorana fermion, these subregion Hilbert spaces are naturally labeled by spin structures, even for a single interval in the ground state. This is in contrast with the twist field formalism, where spin structures only enter in the study of multi-interval entanglement \cite{coser2016spin} or in thermal states with periodic boundary conditions, i.e. tori \cite{foligno2023entanglement}. 

Two of the four spin structures on the torus\footnote{The remaining two appear as charged moments in the context of fermion parity resolution, discussed below.} are shown to be naturally associated with subregion Hilbert spaces $\cH_A$ in the boundary state formalism. These spin structures are distinguished by the presence or absence of a Majorana zero mode, which in turn shapes the entanglement spectrum on $\cH_A$. Interpreting entanglement spectra as probes of edge physics in topological phases \cite{PhysRevLett.108.196402,PhysRevLett.101.010504}, this zero mode corresponds precisely to the trapped degree of freedom distinguishing trivial from non-trivial SPT phases \cite{Cho:2016xjw, smith2021boundary}. This raises the natural question of how spin structures more generally affect the entanglement properties encoded in $\cH_A$ and how their signatures can be extracted.

The Rényi entropy is found below to be comparatively insensitive to spin structures, at least at leading order. Across all three classes of states considered—those based on the vacuum, the fermion field, and conformal interfaces—the Rényi entropies behave identically and reproduce the twist field results. In the vacuum state the universal logarithmic scaling is recovered \cite{Calabrese:2009qy}, and for the fermion primary excitation the result reduces to one obtained using the tools in \cite{ibanez2012entanglement}. While this provides a consistency check, it highlights the need to employ other information measures to capture non-trivial features of the entanglement spectrum, including the Majorana zero mode. 

A natural first candidate for probing beyond Rényi entropies is the relative entropy, which depends explicitly on the entanglement Hamiltonian \cite{blanco2013relative}. As a distance measure on the space of mixed density matrices, it distinguishes reduced density matrices (RDMs) on a subregion $A$. In the present context, relative entropy is employed to compare RDMs based on the vacuum and fermion states. The boundary state formalism once again reproduces the result expected from the conventional twist field approach \cite{lashkari2016modular,ruggiero2017relative}, indicating that relative entropy captures information independent of how the CFT Hilbert space is factorized.

A more refined probe of the entanglement spectrum is provided by symmetry-resolved entanglement \cite{Belin:2013uta, Goldstein:2017bua}. Whenever a subsystem hosts a global symmetry, symmetry resolution reveals the microscopic organization of its entanglement spectrum. The method has become well established, with numerous theoretical developments \cite{Xavier:2018kqb,Calabrese:2021wvi,Murciano:2020vgh,Horvath:2020vzs,Chen:2021nma,Chen:2021pls,Zhao:2020qmn,Weisenberger:2021eby,Zhao:2022wnp,Baiguera:2022sao,Bonsignori:2020laa,Cornfeld:2018wbg,bonsignori2019symmetry,Murciano:2019wdl,Fraenkel:2021ijv,Tan:2019axb,Murciano:2022lsw,Azses:2021wav,magan2021proof, capizzi2022symmetry, capizzi2022symmetry2, capizzi2023symmetry} and experimental realizations \cite{Lukin19,Neven:2021igr,Vitale:2021lds}, as well as direct applications to SPT phases \cite{Azses:2022nfl,Azses:2020tdz}. Importantly, symmetry resolution is sensitive to how Hilbert space factorization is implemented \cite{di2023boundary, northe2023entanglement}. An exotic feature emerging in the boundary state approach is complete equipartition between symmetry sectors, meaning that each sector contributes equally to the information content, independently of the UV cutoff.

While symmetry resolution has been studied for excited states in the twist field approach \cite{capizzi2020symmetry}, it has been set up for the boundary state approach in \cite{yan2025symmetry}, though contributions from the boundary states were dropped. In the case of the Majorana fermion said contributions are important, so that symmetry resolution is carried out below fully in the boundary state approach, allowing one to analyze explicitly the impact of the spin structures. 

The Majorana fermion theory possesses a $\Z_2 \times \Z_2$ symmetry, of which only the fermion parity subgroup $\Z_2^F$, generated by $(-1)^F$, remains upon restriction to a subsystem $A$. This symmetry organizes the entanglement spectrum into bosonic and fermionic sectors, and fermion parity resolution extracts the information content of each. The subgroup $\Z_2^F$ is a symmetry of both the vacuum and fermion subsystem states. Although the conformal interface itself is not invariant under $\Z_2^F$, the Hilbert space factorization stabilizes the symmetry against the interface, as made precise below. Consequently, all three classes of subsystem states considered—based on the vacuum, fermion, and conformal interfaces—admit fermion parity resolution.

A central result below is that whenever an unpaired Majorana zero mode is present, fermion parity resolution yields complete equipartition between the two symmetry sectors: the information content of bosonic and fermionic states in the entanglement spectrum is equal. This effect, previously observed in the vacuum state and described as symmetry-enforced vanishing of the partition function \cite{Cho:2016xjw}, was linked directly to non-trivial SPT phases. Here it is shown to extend beyond the vacuum, holding also for the fermion excitation and for entire classes of conformal interface states. This feature cannot occur in the Ising model, since there the Majorana zero modes are always paired, if present; see \cite{castro2023two} for recent study. Such behavior had previously been observed in the Kitaev chain \cite{fraenkel2020symmetry}\footnote{I thank Shachar Fraenkel and Moshe Goldstein for pointing me to this reference.} and is expanded upon here in its corresponding CFT. In the absence of the unpaired Majorana zero mode, equipartition is shown below to be broken at an order set by the Ramond sector ground state energy and with a magnitude set by the Ramond charge of the factorization. 

Entanglement has been investigated in the presence of conformal interfaces previously, most prominently on a half-interval terminating on the interface \cite{sakai2008entanglement, Brehm:2015lja, brehm2016entanglement,  gutperle2016note, gutperle2017entanglement}, but see also \cite{roy2022entanglement, roy2024topological} for related work. These studies are conducted without use of the boundary state approach to entanglement. Instead, these works place the replica geometry on a torus allowing the authors to evaluate the action of the interfaces naturally on the bulk CFT Hilbert space. This approach essentially assumes that the Hilbert space of a bulk CFT $\cH$ factorizes into two bulk CFT Hilbert spaces $\cH\to\cH_A^{\textrm{bulk}}\otimes\cH_B^{\textrm{bulk}}$, where at least $\cH_A^{\textrm{bulk}}=\cH$. In consequence, the entanglement spectrum corresponds to the bulk spectrum. 

The analysis is qualitatively different in the boundary state approach, where the subregion $A$ is described by a boundary CFT (BCFT). Unfortunately, it is not known, at least to the author, how to evaluate the action of a conformal interface on a boundary Hilbert space, except for topological defects \cite{kojita2018topological}. Nevertheless, certain universal features can still be accessed, such as the effective central charge controlling the leading behavior of the entanglement entropy \cite{karch2023universality, karch2024universal}. The present work focuses instead on constructing entanglement spectra directly associated with conformal interfaces, rather than considering an interface acting on an existing spectrum as in \cite{sakai2008entanglement, Brehm:2015lja, brehm2016entanglement, gutperle2016note, gutperle2017entanglement}. This requires subsystem states in which the interfaces do not intersect the entangling edges $\partial A$. The resulting states give rise to BCFT Hilbert spaces twisted by the interface, which are constructed below and naturally associated with the entanglement spectrum. These spectra are then analyzed within the framework of fermion parity resolution.

The paper is organized as follows. \Secref{secPreliminaries} reviews the computation of primary-state entanglement in the boundary state approach and summarizes the Majorana fermion BCFT. \Secref{secRenyi} applies these tools to evaluate R\'enyi entropies for the vacuum and fermion states, as well as their relative entropy. \Secref{secSymRes} introduces symmetry resolution and develops fermion parity resolution, the central theme of this work. The vacuum and fermion states are analyzed in this framework. \Secref{secConfInt} reviews conformal interfaces in the Majorana fermion theory, while \Secref{secIntEnt} constructs subsystem states based on conformal interfaces and evaluates their entanglement using the preceding methods. A discussion and outlook are presented in \secref{secOutlook}, and technical details on modular forms are collected in \appref{appModularForms}.

\section{Preliminaries}\label{secPreliminaries}
In this section, all preliminaries required to compute R\'enyi entropies in the Majorana fermion theory are reviewed. Firstly, \secref{secExcitedStates} explains how to compute entanglement of primary excitations \cite{ibanez2012entanglement} in the boundary state approach by adapting and slightly extending the presentation in \cite{yan2025symmetry}. Furthermore, \secref{secFermionBCFT} introduces all aspects of the Majorana fermion BCFT required to evaluate primary state R\'enyi entropies.   
\subsection{Primary State Entanglement in the Boundary State Approach}\label{secExcitedStates}
Within the boundary state approach to entanglement between spatial regions $A$ and $B$, factorizations of Hilbert space are treated as mappings
\begin{equation}\label{iota_def}
 \iota_{\alpha\beta}: \cH \to \cH_{\ab}^A\otimes\cH_{\alpha\beta}^B,
 \qquad
 \iota_\ab:\ket{\phi}\mapsto\iota_\ab\ket{\phi}\,,
 \qquad
 \ket{\phi}\in\cH.
\end{equation}
$\alpha,\beta$ label boundary conditions for the quantum fields to be imposed at two disks of size $\epsilon$ surrounding the entangling edges \cite{Ohmori_2015}. $\epsilon$ represents a UV cutoff, which is sent to zero at the end of most analyses. In this work, $\alpha,\,\beta$ are conformal boundary conditions. For concreteness, the interval $A$ is placed on a radial arc with unit radius in the complex plane, parameterized by $z$, terminating on $a=e^{-\iu \pi R}$ and $b=e^{\iu \pi R}$ with $R\in[0,1/2]$. 

Adapting the setup in \cite{yan2025symmetry} to radial quantization, a primary state $\ket{\phi}=\lim_{z,\bz\to0}\phi(z,\bz)\ket{0}\in\cH$ with weights $h,\bh$ is assigned a cut sphere with two open disks at the entangling edges as RDM, 
\begin{equation}\label{cutSphere}
 \rho_A^\phi
 =
 \tr_{B}[\iota_{\alpha\beta}\ketbra{\phi}{\phi}\iota_{\alpha\beta}]
 =
 \frac{1}{N}
 \raisebox{-.45\height}{\includegraphics[scale=.22]{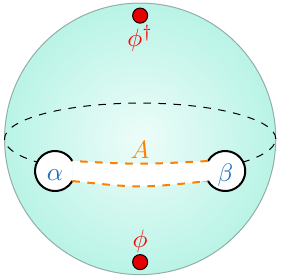}}
\end{equation}
where $\bra{\phi}=\lim_{z,\bz\to\infty}z^{2h}\bz^{2\bh}\bra{0}\phi^\dagger(z,\bz)$ \footnote{Representation labels can be assigned to the fields so that the notation would become $\phi\to\phi_{i\bi}$ and $\phi^\dagger\to\phi_{i^+\bi^+}$. In the present work, only selfconjugate representations appear, $i^+=i$, rendering such a distinction superfluous.} and $N$ normalizes $\rho_A^\phi$. The ket $\ket{0}$ is taken to be the vacuum in holomorphic and antiholomorphic sector. This configuration is mapped onto a cut annulus via
\begin{equation}
 \xi(z)
 =
 e^{-\iu \pi R}\frac{e^{\iu\pi R}-z}{z-e^{-\iu \pi R}}
\end{equation}
The point $b$ is mapped to the origin and a boundary state $\bket{\beta}$ is imposed on its surrounding disk, while the point $a$ is mapped to infinity on the Riemann sphere and the boundary state $\bbra{\alpha}$ is imposed on its surrounding disk. The field insertions are mapped to $\xi(\infty)=e^{\iu\pi(1-R)}$, $\xi(0)=e^{\iu\pi(1+R)}$. The phase $e^{-\iu R}$ ensures that the interval $A$ is mapped onto the positive real axis of the $\xi$-plane. 

\begin{figure}
\begin{center}
 \includegraphics[scale=.25]{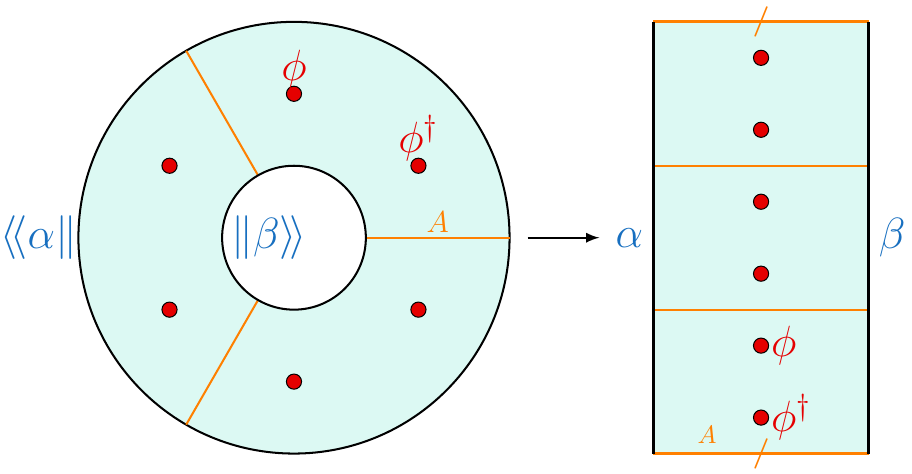}
 \end{center}
 \caption{Left: $n=3$ copies of \eqref{cutSphere} are glued cyclically and mapped onto an annulus by the uniformization map $\xi^{1/n}$. The excized disks are mapped to the inner and outer circle. The boundaries are no longer open and boundary states are imposed on the disks. The interval loci (orange) split the annulus into three equivalent regions. Right: Conformal equivalence with a cylinder slab coordinatized by either \eqref{twCoordn} or \eqref{wCoordn}.}
 \label{figEntMap}
\end{figure}

In order to compute $\tr_\ab[(\rho_{A}^\phi)^n]$, $n$ cyclically glued copies of the sphere \eqref{cutSphere}, each parameterized by $z_m$ with $m=1,\dots,n$, are now mapped onto an annulus via the uniformization map $\xi\to\xi^{1/n}$, see \figref{figEntMap}. The resulting $n$-replica annulus is mapped onto a cylinder slab of circumference 1 via
\begin{equation}\label{twCoordn}
 \tw(z_m)
 =
 \frac{1}{2\pi n \iu}
 \log\xi(z_m)
 =
 \frac{1}{2\pi n\iu}
 \log\xi(z)+\frac{m-1}{n}
\end{equation}
with modular nome $\tq^{1/n}=e^{2\pi \iu\ttau/n}=e^{-2\wi/n}$ and parameter $\ttau=\iu\frac{\wi}{\pi}$, which are expressed in terms of the slab's width
\begin{equation}\label{width}
 \wi
 \equiv
 2\pi|\tw(b(1-\iu\pi\epsilon))-\tw(a(1+\iu\pi\epsilon))|
 =
 2\log\biggl(\frac{2}{\pi\epsilon}\sin(\pi R)\biggr)\,.
\end{equation}
The field insertion points are 
\begin{equation}\label{InsertionsTwN}
  \tw(\infty_m)
 =
 \frac{1-R}{2 n}+\frac{m-1}{n},
 \quad
 \tw(0_m)
 =
 \frac{1+R}{2 n}+\frac{m-1}{n}
\end{equation}
As boundary states are constructed from elements in $\cH$\footnote{Boundary states are non-normalizable in $\cH$ and thus strictly speaking not elements of $\cH$.}, the description is fully in terms of bulk degrees of freedom. Up until here, all is in analogy with \cite{yan2025symmetry}. It shall be convenient in this work to perform furthermore a modular $\modS$ transformation, since the entanglement spectrum is in one-to-one correspondence with the boundary spectrum \cite{cardy2016entanglement, di2023boundary, northe2023entanglement}. $\modS$ is implemented by $\tau=-1/\ttau=\iu\frac{\pi}{\wi}$ ($q=e^{-2\pi^2/\wi}$) and a rescaling,
\begin{align}\label{wCoordn}
 w(z_m)
 =
 n\tau\,\tw(z_m)
 =
 \frac{1}{2\wi}\log\xi(z_m)
\end{align}
The modular nome is now $q^n$ and the field insertions are mapped to
\begin{equation}\label{wCoordnInsertions}
 w(\infty_m)
 =
 \iu\frac{\pi}{2\wi}(1-R+2(m-1))\,,
 \qquad
 w(0_m)
 =
 \iu\frac{\pi}{2\wi}(1+R+2(m-1))\,.
\end{equation} 
While the $\tw$ frame is best for calculations since $\tq\to0$ when $\epsilon\to0$, the frame $w$ is best for (representation theoretic) analysis of the entanglement spectrum, see for instance \cite{cardy2016entanglement, di2023boundary, northe2023entanglement}. The right-hand side of \figref{figEntMap} shows the final configuration. 

A perk of the $w$ coordinates \eqref{wCoordn} is that the Jacobian transformation of the primary $\phi$ with holomorphic dimension $h$ drops out of the RDM. Indeed, the Jacobian is independent of the replica parameters $m$ and $n$,
\begin{equation}
 \frac{\p w(z_m)}{\p z_m}
 =
 \frac{\p w(z)}{\p z}
\end{equation}
where on the right-hand side $z$ is to be understood as for a single copy, i.e. $n=1$. In particular, 
\begin{align}
 \left(z_m^2\frac{\p w(z_m)}{\p z_m}\right)^h\biggl|_{z_m=\infty_m}
 =
 \left(\frac{\p w(z_m)}{\p z_m}\right)^h\biggl|_{z_m=0_m}
 =
 \left(\frac{\tau}{\pi}\, \sin(\pi R)\right)^h
\end{align}
Because the Jacobian is a finite multiplicative factor independent of $n$, it cancels out of the RDM $\rho_A^\phi$ by virtue of the normalization $\tr\rho_A^\phi=1$. The argument applies identically to the anti-holomorphic sector.

In conclusion, the RDM \eqref{cutSphere} takes the useful form
\begin{equation}\label{RDM}
 \rho_{\ab}^{\phi}
 =
 \frac{1}{Z^{\phi}_{\ab}(q)}
 \raisebox{-.5\height}{\includegraphics[scale=.12]{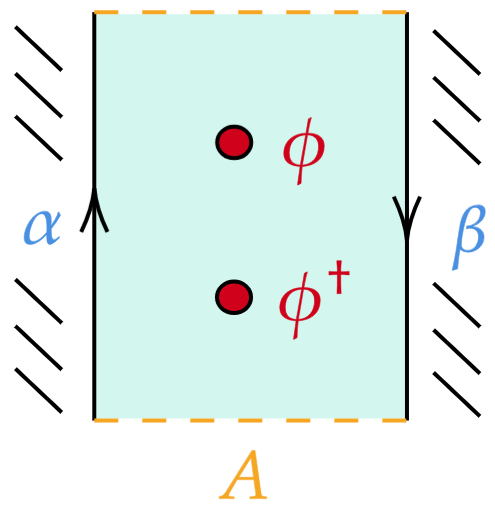}}\,,
 \qquad
 Z^{\phi}_{\ab}(q)
 =
 \tr_{\ab}\left[q^{H_{\ab}} \phi(w(0))\phi^\dagger(w(\infty))
 \right]
\end{equation}
The strip has height $\tau$, width $1/2$ and is understood to carry the evolution operator $q^{H_\ab}$, where $H_\ab$ is the Hamiltonian on the strip, and evolves from bottom to top. Hence for the vacuum, $\phi=\id$, the expected result $\rho_\ab^\id=q^{H_\ab}/Z_\ab^\id(q)$ is recovered \cite{cardy2016entanglement}. Instead of labeling the RDM by the interval $A$, as is common, the boundary labels $\ab$ are dressing it here. In the following, the state $\phi$ is referred to as \textit{global state}, in contrast to the subsystem states $\rho_\ab^\phi$. The frame \eqref{RDM} is most useful in setting up calculations in the following.

\subsection{Majorana Fermion BCFT}\label{secFermionBCFT}
The theory of interest is the free fermion CFT with action
\begin{equation}\label{action}
 S=\int d^2z(\psi\bp\psi-\bpsi\p\bpsi)
\end{equation}
Its bulk CFT is discussed in standard introductory texts \cite{Ginsparg, DiFrancesco:1997nk}, so that emphasis falls on its BCFT here, see \cite{Recknagel:2013uja, gaberdiel2002d, northe2025young} for reviews. On the plane, fermions acquire a sign when encircling the origin, $\psi(e^{2\pi\iu}z)=\sigma\psi(z)$, where $\sigma=1$ in the Neveau-Schwarz (NS) sector and $\sigma=-1$ in the Ramond (R) sector. When placed on the upper half-plane, the boundary conditions
\begin{align}
 \bpsi(\bz)=\alpha\psi(z)|_{\bz=z>0},
 \qquad 
 \bpsi(\bz)=\sigma\alpha\psi(z)|_{\bz=z<0}
\end{align}
preserve conformal symmetry with $\alpha=\pm1$\footnote{$\alpha=1$ corresponds to Neumann and $\alpha=-1$ to Dirichlet boundary conditions. This terminology is not employed in this work however.}. For now, the exposition is restricted to the NS sector, since the fermion $\ket{\psi}$ and vacuum state $\ket{0}$, whose entanglement is investigated below, naturally fall into it. The R sector is discussed in \secref{secRoleRbkets} in the context of charged moments. The NS sector boundary states are 
\begin{align}\label{bkets}
 \bket{\alpha}
 =
 \exp\left(\iu\alpha\sum_{s\in\N_0+1/2}\psi_{-s}\bpsi_{-s}\right)\ket{0}\,,
 \qquad
 (\psi_s-\iu\alpha\bpsi_{-s})\bket{\alpha}=0\,,
\end{align}
where $\psi_s$ are fermionic modes in the NS sector, $\psi(z)=\sum_{s\in\Z+1/2}\psi_sz^{-s-1/2}$. Utilizing the bulk Hamiltonian $H_{ns}$ in the NS sector, the overlaps $\bbra{\alpha}\tq^{\frac{H_{ns}}{2}}\bket{\beta}=Z_{\ab}(q)$ yield two inequivalent boundary state sums
\begin{subequations}\label{bdySpec}
\begin{align}
 Z_{\aa}(q)
 &=
  \tr_{NS}[q^{H_{NS}}]
 =
 \sqrt{\frac{\vartheta_3(q)}{ \eta(q)}}
 =
 Z_3(q)\label{bdySpecZ3}\\
 Z_{(-\alpha)\alpha}(q)
 &=
 \frac{1}{\sqrt{2}}\tr_{R}[q^{H_R}]
 =
 \sqrt{\frac{\vartheta_2(q)}{ \eta(q)}}
 =
 Z_2(q)\label{bdySpecZ2}
\end{align}
\end{subequations}
The Hamiltonians $H_\aa=H_{NS}$ and $H_{(-\alpha)\alpha}=H_R$ evolve in time on the strip and are, respectively, in the NS and R sector. Definitions of the Jacobi $\vartheta$ functions are collected in \appref{appModularForms}. Note the $\sqrt{2}$ factor in \eqref{bdySpecZ2}. It is typical of unpaired Majorana zero modes, and the Cardy constraint must be relaxed in fermionic models to accomodate this possibility \cite{smith2021boundary}. Indeed, expanding out $Z_2(q)$ in $q$ one finds an overall factor of $\sqrt{2}$, which is conveniently absorbed in the trace, i.e. $\tr_{(-\alpha)\alpha}=\frac{1}{\sqrt{2}}\tr_R$. In contrast $\tr_{\alpha\alpha}=\tr_{NS}$. The resulting partition functions $Z_\nu$ correspond to well-known chiral spin structures on the torus: for $\nu=3$ the fermion is antiperiodic on both cycles and for $\nu=2$ it is (anti-)periodic in the (temporal) spatial direction \cite{DiFrancesco:1997nk}. 

Given a spin structure $\nu$, the fermion propagator at modular nome $q$ is expressed in terms of Jacobi $\vartheta$ functions \cite{DiFrancesco:1997nk}, 
\begin{align}\label{Propagator}
 \corr{\psi(w)\psi(v)}_\nu(q)
 =
 \frac{\vartheta_\nu(w-v,\tau)2\pi\eta^3(q)}{\vartheta_\nu(0,\tau)\vartheta_1(w-v,\tau)}
\end{align}
See \eqref{ChargedJacobiTheta} for their definition. Their modular $\modS$ transformations from modular coordinates $(q^n,w)$ to $(\tq^{1/n},\tw)$ are important below, 
\begin{align}\label{PropagatorModS}
 \corr{\psi(w)\psi(v)}_\nu(q^n)
 =
 \frac{M_{\nu\mu}}{n\tau}
 \corr{\psi(\tw)\psi(\tilde{v})}_\mu(\tq^{1/n})
\end{align}
where $M_{24}=M_{42}=M_{33}=1$ and the remaining $M_{\nu\mu}$ vanish. In this text, the modular nomes $q^n,\,\tq^{1/n}$ are used to keep track of the replica geometry, i.e. \eqref{Propagator} has $n=1$ replica, while \eqref{PropagatorModS} has $n$ replica. Finally, when the UV cutoff $\epsilon$  shrinks away ($\tq\to0$), the propagator on the plane is recovered,
\begin{align}
 \lim_{\tq\to0}\corr{\psi(\tw)\psi(0)}_{3,4}(\tq^{1/n})\label{Propagator34}
 &= 
 \frac{\pi}{\sin(\pi\tw)}
\end{align}

\section{Subsystem Entropies for Global Primary States}\label{secRenyi}

Entanglement between spatial regions $A$ and $B$ is quantified for pure states $\ket{\phi}$ by the R\'enyi entropies 
\begin{equation}
 S_n(\rho_\ab^\phi)
 =
 \frac{1}{(1-n)}\log\tr[(\rho_\ab^\phi)^n]
 =
 \frac{1}{1-n}\log\left[\frac{Z_{\ab}^\phi(q^n)}{(Z_{\ab}^\phi(q))^n}\right] ,
\end{equation}
where $Z_\ab^\phi(q)$ is as in \eqref{RDM}, whereas $Z_\ab^\phi(q^n)$ covers $n$ replica sheets and is thus understood to carry $n$ insertions of $\phi\,\phi^\dagger$. The R\'enyi entropy is evaluated for the global vacuum $\ket{0}$ with $h=\bh=0$ in \secref{secRenyiVacuum} and for the fermion state $\ket{\psi}$ with $h=1/2$, $\bh=0$ in \secref{secRenyiPsi}. Moreover, one relative entropy between $\rho_\ab^\id$ and $\rho_\ab^\psi$ is computed in \secref{secRelEntropies}.

\subsection{R\'enyi Entropies for the Global Vacuum State}\label{secRenyiVacuum}
Beginning with the vacuum $\ket{0}$, $\phi=\id$ must be set and the entire entanglement is due to the factorization \eqref{iota_def}. The two inequivalent choices $\iota_\aa$ and $\iota_{(-\alpha)\alpha}$ lead to entanglement spectra described by the spin structures \eqref{bdySpec}, whose R\'enyi entropies are
\begin{align}\label{vacuumRenyis}
 S_n(\rho_\ab^\id)
 &=
 \frac{1}{1-n}\log\left[M_{\nu\mu}\frac{Z_{\mu}(\tq^{1/n})}{(Z_{\mu}(\tq))^n}\right]
 \overset{\tq\to0}{=}
 \frac{\wi}{24}\frac{n+1}{n}+\cO(\epsilon^1)
 \,,
\end{align}
where $Z_{\ab}^\id(q^n)=Z_{\nu}(q^n)$ has been identified, with spin structure $\nu=3$ for $\beta=\alpha$ and $\nu=2$ for $\beta=-\alpha$, see \eqref{bdySpec} and modular $\modS$ transformed. For $\tq\to0$, the celebrated logarithmic scaling \eqref{width} is reproduced \cite{Calabrese:2009qy}. Note that $\cO(\epsilon^0)$ is absent since the g-factors $\log\gf_\alpha=\log\langle0\bket{\alpha}=0$ for \eqref{bkets}. Therefore the leading orders of the R\'enyi entropies do not distinguish the two factorizations in the case at hand.

Nevertheless, the information content of $\rho_{(\pm\alpha)\alpha}^\id$ is very distinct, as seen in multiple ways. Clearly, the $S_n(\rho_{(\pm \alpha)\alpha}^\id)$ differ for some order of $\epsilon$ given their distinct spin structures,
\begin{equation}\label{BoundaryZs}
 Z_\aa^\id(q)=Z_3(q)=\chi_0(q)+\chi_{1/2}(q),
 \qquad
 Z_{(-\alpha)\alpha}^\id(q)=Z_2(q)=\sqrt{2}\chi_{1/16}(q)
\end{equation}
where $\chi_i$ are Virasoro characters of weight $i=0,\,1/2,\,1/16$. Hence, the spectra of the vacuum's entanglement Hamiltonian $K_\ab^\id=-\frac{1}{2\pi}\log \rho_\ab^\id$ \cite{Roy_2020},
\begin{equation}\label{ESvacuum}
 \varepsilon_\ab^\id(i,k)=\frac{\pi}{\wi}\left(h_i+k-\frac{\cc}{24}\right)+\frac{1}{2\pi}\log Z_\ab^\id ,\qquad k\in\N
\end{equation}
are different. Descendants are counted by $k\in \N$. Note in particular the $\sqrt{2}$-fold degeneracy in the spectrum $Z_{(-\alpha)\alpha}^\id(q)$. This factor is the partition function $Z_{\psi_0}$ of an unpaired Majorana zero mode \cite{smith2021boundary} satisfying
\begin{equation}
 \psi_0^\dagger=\psi_0,
 \qquad
 \{\psi_0,\psi_0\}=1\,,
 \qquad
 Z_{\psi_0}=\sqrt{2}\,,
\end{equation}
which plays a central role in this paper. As seen by employing \eqref{SpinStructures} in the $q$-channel, $\psi_0$ appears when comparing the (infinte) ranks of the RDMs $\rho_\ab^\id$ \footnote{I thank Pedro J. Martinez for suggesting this check.},
\begin{equation}
 \lim_{n\to0}\left(S_n(\rho_{-\alpha\alpha}^\id)-S_n(\rho_\aa^\id)\right)
 =
 \frac{1}{2}\log2
 =
 \log Z_{\psi_0}\,.
\end{equation}
Moreover, the Majorana zero mode contributes to the vanishing of the constant order of $S_n(\rho_{(-\alpha)\alpha}^\id)$, i.e. the g-factor. This is seen by writing the R\'enyi entropy in terms of the rightmost expression in \eqref{BoundaryZs},
\begin{align}
 S_n(\rho_{(-\alpha)\alpha}^\id)
 &=
 \frac{1}{1-n}\log\left[\frac{Z_{\psi_0}\,\chi_{1/16}(q^n)}{(Z_{\psi_0}\chi_{1/16}(q))^n}\right]\notag\\
 &\overset{\tq\to0}{\simeq}
 \frac{1}{1-n}\log\left[\frac{\chi_0(\tq^{1/n})}{(\chi_0(\tq))^n}\right]+\log\left(Z_{\psi_0}\modS_{\frac{1}{16},0}\right)
 \simeq
 \frac{\wi}{24}\frac{n+1}{n}+\cO(\epsilon^1)
\end{align}
The second line is due to modular $\modS$ transformation and uses that in the asymptotic regime $\tq\to0$ the vacuum character dominates and more precisely only its vacuum state leading to the final expression. More importantly, the Majorana zero mode contribution cancels the modular matrix entry $\modS_{0\frac{1}{16}}=2^{-1/2}$. Thus, in absence of $\psi_0$ this R\'enyi entropy would feature a term at $\cO(\epsilon^0)$.

\subsection{R\'enyi Entropies for the Global Fermion State}\label{secRenyiPsi}
Turning to the global fermion state $\ket{\psi}=\psi(0)\ket{0}$, the RDM $\rho^\psi_\ab$ is given by \eqref{RDM} with $\phi=\psi$ and $Z^\psi_\ab=Z_\nu(q)\corr{\psi(w)\psi(v)}_\nu(q)$, where $\tr[O\,q^{H}]=\tr[q^H]\corr{O}$ has been employed. The required moments are simplified by
\begin{align}\label{psiMoment}
 \tr_{\ab}\left[(\rho_{\ab}^\psi)^n\right]
 &=
 \frac{\tr_{\ab}\left[q^{nH_{\ab}}\prod_{m=1}^n\psi(w(0_m))\psi(w(\infty_m))\right]}{(Z_\ab^\psi(q))^n}\notag\\
 &=\frac{1}{(Z_\ab^\psi(q))^n}Z_\nu(q^n)\,\corr{\prod_{m=1}^n\psi(w(0_m))\psi(w(\infty_m))}_\nu(q^n)\notag\\
 %
 %
 &=\tr_{\ab}\left[(\rho_{\ab}^\id)^n\right]\,\frac{\Pf[\corr{\psi(w_k)\psi(w_l)}_\nu(q^n)]}{(\corr{\psi(w_0)\psi(w_\infty)}_\nu(q))^n}
\end{align}
where $Z_\nu=Z_\ab^\id$ has been employed in the last step. For $n>1$, all field insertions  are now labeled by $k,l$, i.e. $w_k$ stands for either $w(0_m)$ or $w(\infty_m)$ with $m=1,\dots,n$, as in \eqref{wCoordnInsertions}. For $n=1$, the Pfaffian reduces to the two-point function with insertions at $w_\infty=w(\infty)$ and $w_0=w(0)$. 

R\'enyi entropies at finite UV cutoff are easily read off from \eqref{psiMoment} after use of \eqref{PropagatorModS} for any $n\neq1$. At zero UV cutoff the Pfaffian can be further manipulated as follows
\begin{align}\label{diffsn}
 \diffsn{n}
 &:=
 \lim_{\tq\to0}\,\bigl(S_n(\rho_{\ab}^\psi)-S_n(\rho_{\ab}^\id)\bigr)\\
 &=
 \lim_{\tq\to0}\frac{1}{1-n}\log\left[\frac{M_{\nu\mu}}{n^n}\frac{\Pf[\corr{\psi(\tw_k)\psi(\tw_l)}_\mu(\tq^{1/n})]}{(\corr{\psi(\tw(0))\psi(\tw(\infty))}_\mu(\tq))^n}\right]\notag\\
 &= 
 \frac{1}{1-n}\log\left[\frac{1}{n^n}\frac{\Pf[\sin(\pi(\tw_k-\tw_l))^{-1}]}{(\sin(\pi R))^{-n}}\right]\,,\notag
\end{align}
where \eqref{PropagatorModS} and \eqref{Propagator34} have been used. Because the Pfaffian sums over products of pairwise contractions and each contraction has a well-defined limit, the limit $\tq\to0$ may be drawn into the Pfaffian\footnote{This is true at least for $n>1$. For $n=1$, order of limits issues may arise, which can only be dispensed by analytically continuing the second line in \eqref{diffsn} in $n$. Such a continuation is not known to the author so that strictly speaking validity of \eqref{diffsn} for $n=1$ is assumed.}, leading to the final line.  Note that $\diffsn{n}$ coincides for $\iota_{\aa}$ and $\iota_{(-\alpha)\alpha}$ and that furthermore the calculation reduces to one performed in the conventional twist field formalism presented in \cite{ibanez2012entanglement}. 

The Pfaffian is evaluated by $\Pf^2[\sin(\pi(\tw_k-\tw_l))^{-1}] = \det\mathbb{H}$, where
\begin{align}
 \mathbb{H}_{jk}
 =
 \begin{cases}
  \sin(\pi(\tw_j-\tw_k))^{-1},&\, \textrm{if } j\neq k\\
  0 &\, \textrm{if } j=k
 \end{cases}
 \quad 
\textrm{ and }
\quad
 \tw_j
 =
 \begin{cases}
  \tw_{(n)}(0_m),&\, \textrm{if } j\leq n\\
  \tw_{(n)}(\infty_m),&\, \textrm{if } j>n\\
 \end{cases}
\end{align}
The square needed in relating the Pfaffian to the determinant is crucial in distinguishing the fermionic case from the bosonic case in \cite{calabrese2014entanglement}, where precisely this determinant has been brought into the following form,
\begin{equation}
 \det\mathbb{H}
 =
 4^n\frac{\Gamma^2\left(\frac{1+n}{2}+\frac{n}{2\sin(\pi R)}\right)}{\Gamma^2\left(\frac{1-n}{2}+\frac{n}{2\sin(\pi R)}\right)}
\end{equation}
Plugging this into the difference of entanglement entropies yields,
\begin{align}\label{diffsnResult}
 \diffsn{n}
 &=
 \frac{1}{1-n}\log\left[\left(\frac{2\sin(\pi R)}{n}\right)^{n}\frac{\Gamma\left(\frac{1+n}{2}+\frac{n}{2\sin(\pi R)}\right)}{\Gamma\left(\frac{1-n}{2}+\frac{n}{2\sin(\pi R)}\right)}\right]
\end{align}
Comparing with analogous results on the free boson theory, it is readily seen that \eqref{diffsnResult} is one half smaller than its bosonic analog computed in  \cite{ruggiero2017relative}, where $\psi\to\p\phi$. Such behavior is traced back to the difference in conformal weights, $h_\psi=h_{\p\phi}/2$ and the fact that both are free fields. In conclusion, the R\'enyi entropies are expanded for small UV cutoff as
\begin{align}\label{RenyiPsi}
 S_n(\rho^\psi_\ab)=\frac{\wi}{24}\frac{n+1}{n}+\diffsn{n}+\dots
\end{align}
For the entanglement entropy one easily finds from \eqref{diffsnResult},
\begin{equation}\label{diffsn1}
 \diffsn{1}
 =
 -\log[2\sin(\pi R)]-\sin(\pi R)-\Psi\left(\frac{1}{2\sin(\pi R)}\right)
\end{equation}
where $\Psi(z)=\Gamma'(z)/\Gamma(z)$ is the digamma function. This term also corrects R\'enyi entropies in the quarter filled Hubbard model \cite{calabrese2014entanglement}. In closing this subsection, it is remarked that it is not clear what the entanglement spectrum for $\rho_\ab^\psi$ is, in contrast to the vacuum RDM $\rho_\ab^\id$, see \eqref{ESvacuum}. 

\subsection{Relative Entropy}
\label{secRelEntropies}
Now that two types of RDMs, one for $\phi=\id$ and another one for $\phi=\psi$, are at hand, they can be compared. 
Two density matrices $\rho,\sigma\in\End(\cH^A)$ can be distinguished by means of the relative entropy \cite{lashkari2016modular}
\begin{equation}
 S(\rho||\sigma)
 =
 \lim_{n\to1}\frac{1}{1-n}\log\left[\frac{\tr_A\bigl(\rho\sigma^{n-1}\bigr)}{\tr_A(\rho^n)}\right]
\end{equation}
The relative entropy measures how well a density matrix $\sigma$ approximates the true state of a system $\rho$. $S(\rho||\sigma)$ enjoys some appealing properties. It is non-negative, $S(\rho||\sigma)\geq0$, and increases under inclusion, that is if the interval $A$ is enlarged, then $S(\rho||\sigma)$ grows. While the relative entropy is not symmetric, $S(\rho||\sigma)\neq S(\sigma||\rho)$, and has therefore no chance of being a metric, it still provides a useful distance measure on the space of density matrices. Most importantly, it is UV finite, rendering it a useful tool in proving mathematically rigorous statements in the quantum information theory of quantum fields \cite{casini2016g, casini2023entropic, casini2022lectures}.

In the Majorana fermion theory, it is interesting to ask how well the vacuum state $\rho_\ab^\id$ approximates the fermion state $\rho_\ab^\psi$, because the former is more easily accessible than the latter. The ratio of traces is evaluated by elementary manipulations
\begin{align}
 \frac{\tr_\ab[\rho_\ab^\psi(\rho_\ab^\id)^{n-1}]}{\tr_\ab[(\rho_\ab^\psi)^n]}
 &=
 \frac{\tr_\ab[q^{nH_\ab}\psi(w(0_n))\psi(w(\infty_n))]}{Z_\ab^\psi(q)(Z^\id_\ab(q))^{n-1}\,\tr_\ab[(\rho_\ab^\psi)^n]}\notag\\
 &=
 \frac{\corr{\psi(w(0_n))\psi(w(\infty_n))}_\nu(q^n)}{\corr{\psi(w_0)\psi(w_\infty)}_\nu(q)}\,
 \frac{(\corr{\psi(w_0)\psi(w_\infty)}_\nu(q))^n}{\Pf[\corr{\psi(w_k)\psi(w_l)}_\nu(q^n)]}
\end{align}
The first line uses \eqref{RDM} for $\phi=\id$ and $\phi=\psi$, while the second uses \eqref{psiMoment}. The second ratio leads naturally to differences of Renyi entropies, and thus in the relative entropy one finds
\begin{align}\label{RelEntropy}
 S(\rho_\ab^\psi||\rho_\ab^\id)
 &=
 \lim_{\tq\to0}\lim_{n\to1}
 \frac{1}{1-n}\log\left(
 \frac{\corr{\psi(w_{(n)}(0))\psi(w_{(n)}(\infty))}_\nu(q^n)}
 {\corr{\psi(w_0)\psi(w_\infty)}_\nu(q)}
 \right)
 -
 \diffsn{1}\notag\\
 &=
 1-\lim_{\tq\to0}\lim_{n\to1}M_{\nu\mu}
 \left(
 \frac{\p_n\corr{\psi(\tw_{(n)}(0))\psi(\tw_{(n)}(\infty))}_\mu(\tq^{1/n})}
 {\corr{\psi(\tw_{(n)}(0))\psi(\tw_{(n)}(\infty))}_\mu(\tq^{1/n})}
 \right)
 -
 \diffsn{1}\notag\\
 &=
 1-\lim_{n\to1}
 \left(
 \sin(\pi(\tw_{(n)}(0)-\tw_{(n)}(\infty)))\p_n\frac{1}
 {\sin(\pi(\tw_{(n)}(0)-\tw_{(n)}(\infty)))}
 \right)
 -
 \diffsn{1}\notag\\
 &=
 1-\pi R\cot(\pi R)+\log[2\sin(\pi R)]+\sin(\pi R)+\Psi\left(\frac{1}{2\sin(\pi R)}\right)
\end{align}
where in the first line \eqref{RDM} has been used with $\phi=\psi,\,\phi=\id$. The second line modular $\modS$ transforms \eqref{PropagatorModS} and employs l'Hospitals rule. The third line exhanges the limits and uses \eqref{Propagator34}. In the first summand, it can be checked by tedious calcuation that the limits indeed commute, while $\lim_{\tq\to0}\lim_{n\to1}\diffsn{n}=\lim_{n\to1}\lim_{\tq\to0}\diffsn{n}$ is an assumption. The last line uses $\tw_{(n)}(0)-\tw_{(n)}(\infty)=R/n$, see \eqref{InsertionsTwN} and \eqref{diffsn1} is plugged in.

Observe that the limit $\tq\to0$ removes all dependence on the boundary conditions $\alpha,\,\beta$ indicating that the relative entropy is insensitive to the factorization \eqref{iota_def} when comparing the RDMs $\rho_\ab^{\id}$ and $\rho_\ab^{\psi}$. As required, the relative entropy is monotonically increasing with the interval size $R$. 

Comparing with similar results on the free boson current $\p\varphi$ in \cite{ruggiero2017relative}, one immediately sees that the fermionic result here is exactly half as large as its bosonic cousin. Since the two-point  function of free fields is the main ingredient in this calculation, it is hardly surprising. After all this correlator is controlled by the conformal dimension and they differ precisely by this factor $h_{\p\varphi}=2h_\psi=1$. One can therefore say that the vacuum subsystem state $\rho_\ab^\id$ approximates the fermion state $\rho_\ab^\psi$ better than it could approximate the bosonic current subsystem state $\rho_\ab^{\p\varphi}$. Though not proven by the present calculation, it is likely that $\rho_\ab^\id$ approximates a state $\rho_\ab^\phi$ better the closer $h_\phi$ is to $h_\id=0$. The inverse question, i.e. how well $\rho_\ab^\psi$ approximates $\rho_\ab^\id$, is more difficult to address for lack of an appropriate analytic continuation \cite{ruggiero2017relative}. 

So far the impact of the factorization \eqref{iota_def} on entropic functions \eqref{vacuumRenyis}, \eqref{RenyiPsi} and \eqref{RelEntropy} appears to be rather limited. To fully appreciate the physical information carried by the entangling edges and the factorization \eqref{iota_def}, the entanglement spectrum needs to be contemplated with care, as done in the following.

\section{Fermion Parity Resolution of Entanglement}\label{secSymRes}
The spectrum of fermionic theories contains bosonic and fermionic states, which are mapped into each other by any fermionic mode. To obtain a deeper understanding of the entanglement spectrum, the present section asks which fermion parity sector contains more information. As shown here, presence of the Majorana zero mode forces bosonic and fermionic sectors to have an equal information count. \Secref{secSubsystemSym} discusses the symmetries of the fermion theory before and after Hilbert space factorization. Furthermore, the definition of symmetry-resolved entropy is provided. In \secref{secParityRes}, these tools are applied to the vacuum and fermion state. \Secref{secRoleRbkets} discusses the  connection between Ramond sector boundary states and charged moments and demonstrates how they quantify the breaking of equipartition.  

\subsection{Subsystem Fermion Parity and Symmetry Resolution}\label{secSubsystemSym}

The fermion theory \eqref{action} has a global $G=\Z_2^{(F)}\times\Z_2^{(F_L)}$ symmetry generated by
\begin{align}
 (-1)^F:\quad \psi\to-\psi, \quad \bpsi\to-\bpsi, 
 \qquad
 (-1)^{F_L}:\quad \psi\to-\psi, \quad \bpsi\to\bpsi
\end{align}
$G$ is completed by the group unit $e$ and the right-moving chirality operator $(-1)^{F_R}$.

Before tracing over $B$, the density matrices $\ketbra{0}{0}$ and $\ketbra{\psi}{\psi}$ clearly commute with $G$. However, the factorization $\iota_\ab$ may itself break symmetries \cite{di2023boundary}. To understand if a factorization preserves a subgroup $G'\subseteq G$, it suffices to check whether all grouplike defects in $G'$ can end topologically on the two boundaries simultaneously. This secures that the entanglement Hamiltonian $K_\ab\propto H_\ab$ of the vacuum state under the factorization $\iota_\ab$ commutes with implementations of $g\in G'$ \cite{choi2023remarks}\footnote{Because the symmetries are implemented by topological defects, the symmetry operations in fact commute with the entire Virasoro algebra of the BCFT, i.e. $[G',L_n^{(H)}]=0$}. They are thus proper symmetries of the BCFT. Diagrammatically, topological endability is phrased as
\begin{equation}\label{TopEndability}
 \raisebox{-.45\height}{\includegraphics[scale=.25]{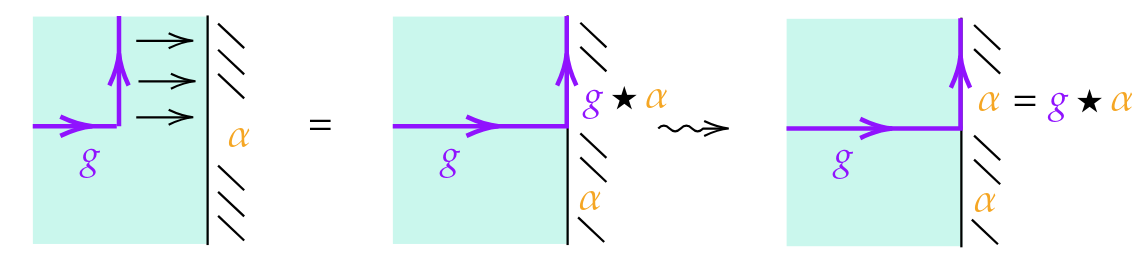}}
\end{equation}
where $\star$ denotes the fusion of the boundary and topological defect $D_g$. Grouplike defects map simple boundary conditions, as the ones in \eqref{bkets}, into simple boundary conditions. For the defect line attaching to the boundary to be able to move up and down, the junction field between defect and boundary must have conformal dimensions $h=\bh=0$, which happens if $g\star\alpha=\alpha$. 
A simple diagnostic to check topological endability on the boundary for grouplike defects is therefore $\bket{g\star\alpha}=D_g\bket{\alpha}=\bket{\alpha}$ for all $g\in G'$. 

The action of $G$ on the boundary states \eqref{bkets} is
\begin{align}\label{GonBket}
 (-1)^F\bket{\alpha}=\bket{\alpha},
 \qquad
 (-1)^{F_L}\bket{\alpha}=\bket{-\alpha}
\end{align}
where $(-1)^F\ket{0}=(-1)^{F_L}\ket{0}=\ket{0}$ was used. Hence only the subgroup $G'=\Z_2=\{e,(-1)^F\}$ is a proper symmetry on the subsystem $\cH^A_{\ab}$. Diagrammatically, it is easily seen that injecting a field $\phi=\id,\,\psi$ does not hinder $G'$ from being the subsystem's symmetry. Indeed, $\rho_\ab^\phi(-1)^F=(-1)^F\rho_\ab^\phi$ is seen as follows
\begin{equation}\label{FermionNumberAction}
 \raisebox{-.45\height}{\includegraphics[scale=.25]{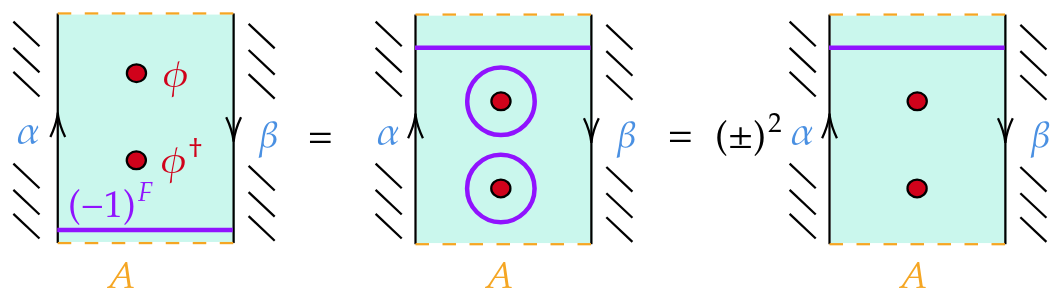}}
\end{equation}
where each encircled field amounts to either $(-1)^F\ket{\phi=\id}=+\ket{\id}$ or $(-1)^F\ket{\phi=\psi}=-\ket{\psi}$ \cite{frohlich2007duality, northe2025young}.
The $\Z_2$ symmetry organizes the subsystem's Hilbert space into subspaces with fermion number $(-1)^F=1$ ($a=+$) and $(-1)^F=-1$ ($a=-$),
\begin{equation}\label{SymmetrySplitting}
 \cH_\ab^A=\bigoplus_{a=\pm}V_a\otimes \cH_\ab^a
\end{equation}
where $V_a$ are irreducible representations of $\Z_2$ and $\cH^a_\ab$ is its multiplicity space. The RDM $\rho_\ab^\phi\in\End(\cH_\ab^A)$ commutes with $(-1)^F$, leading to a block diagonal decomposition,
\begin{align}
 \rho_\ab^\phi
 =
 \bigoplus_{a=\pm}\prob_a^\phi\rho_\ab^\phi(a),
 \qquad 
 \rho_\ab^\phi(a)
 =
 \frac{\rho_\ab^\phi\Pi_a}{\prob^\phi_a}
 =\frac{1}{\prob^\phi_a}
 \raisebox{-.5\height}{\includegraphics[scale=.175]{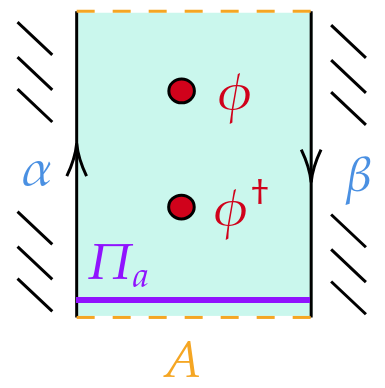}}\,,
\end{align}
via projectors $\Pi_\pm=\frac{1}{2}(e\pm(-1)^F)$ onto $V_\pm$ with group unit element $e$. The probabilities $\prob_a^\phi=\tr_\ab[\Pi_a\rho_\ab^\phi]$ are the $n=1$ case of the $a$-sector moments,
\begin{equation}\label{ProjectedMoments}
 \cZ^\phi_\ab(a|n)
 =
 \tr_\ab\left[\Pi_a(\rho^\phi_\ab)^n\right]\,,
\end{equation}
which in turn give rise to the symmetry-resolved R\'enyi entropies \cite{Goldstein:2017bua},
\begin{align}\label{SREE}
 S_n(\rho_\ab^\phi(a))
 &=
 \frac{1}{1-n}\log\left[\frac{\cZ^\phi_\ab(a|n)}{(\cZ^\phi_\ab(a|1))^n}\right]
\end{align}
When $S_n(\rho_\ab^\phi(a))=S_n(\rho_\ab^\phi(b))$ for $a\neq b$, the two sectors $a,\,b$ are said to be equipartitioned \cite{Xavier:2018kqb} and carry the same information count. Typically, one compares only leading orders in $\epsilon$ of \eqref{SREE}, which is called asymptotic equipartition in \cite{northe2023entanglement}, as opposed to complete or exact equipartition, which occurs when $S_n(\rho_\ab^\phi(a))=S_n(\rho_\ab^\phi(b))$ holds at all orders in $\epsilon$.
\subsection{Fermion Parity-Resolved R\'enyi Entropies for Global Primary States}\label{secParityRes}
In the following, the two states $\phi=\id$ and $\phi=\psi$ are analyzed. Turning first to the invisible field $\phi=\id$ and recalling \eqref{BoundaryZs} as well as the elementary relations \cite{Ginsparg}
\begin{align}
 \tr_{NS}\left[q^{H_{NS}}\right]&=\chi_0(q)+\chi_{1/2}(q)\,,&
 \tr_{NS}\left[(-1)^Fq^{H_{NS}}\right]&=\chi_0(q)-\chi_{1/2}(q)\,,\\
 \tr_{R}\left[q^{H_{R}}\right]&=2\chi_{1/16}(q)\,,&
 \tr_{R}\left[(-1)^Fq^{H_{R}}\right]&=0
\end{align}
where $\chi_0,\,\chi_{1/2},\,\chi_{1/16}$ are Virasoro characters. $\chi_0$ has parity even states while $\chi_{1/2}$ has parity odd states in the factorization $\iota_\aa$, whereas in $\iota_{(-\alpha)\alpha}$ both sectors have the same state content,
\begin{align}
 \cZ^\id_\aa(n|+)=\frac{\chi_0(q^n)}{(Z_\aa^\id(q))^n}\,,
 \qquad
 \cZ^\id_\aa(n|-)=\frac{\chi_{1/2}(q^n)}{(Z_\aa^\id(q))^n}\,,\\
 \cZ^\id_{(-\alpha)\alpha}(n|a)=\frac{1}{\sqrt{2}}\frac{\chi_{1/16}(q^n)}{(Z_{(-\alpha)\alpha}^\id(q))^n}\,,\qquad\qquad
\end{align}
Observe once more the $\sqrt{2}$ factor indicating the presence of the Majorana zero mode in $\iota_{(-\alpha)\alpha}$. Consequently,
\begin{subequations}\label{SRREferm}
\begin{align}
 S_n(\rho^\id_\aa(+))
 &=
 \frac{1}{1-n}\log\left[\frac{\chi_0(q^n)}{(\chi_0(q))^n}\right],\\
 S_n(\rho^\id_\aa(-))
 &=
 \frac{1}{1-n}\log\left[\frac{\chi_{1/2}(q^n)}{(\chi_{1/2}(q))^n}\right]\\
 S_n(\rho^\id_{(-\alpha)\alpha}(a))
 &=
 S_n(\rho^\id_{(-\alpha)\alpha})-\log2
 =
 \frac{1}{1-n}\log\left[\frac{\chi_{1/16}(q^n)}{(\chi_{1/16}(q))^n}\right]-\frac{1}{2}\log2
 \label{SRREpma}
\end{align}
\end{subequations}
It has been shown in \cite{northe2023entanglement} that $S_n(\rho_\aa^\id(\pm))$ are asymptotically equipartitioned, i.e. only to $\cO(\epsilon^0)$. In contrast, the parity sectors of the factorization $\iota_{(-\alpha)\alpha}$ are clearly equipartitioned to all orders. This property is enforced by the Majorana zero mode $\psi_0$, and holds true even when picking the fermion as global state,
\begin{equation}\label{SRREpsiEquipartition}
 S_n(\rho^\psi_{(-\alpha)\alpha}(a))
 =
 S_n(\rho^\psi_{(-\alpha)\alpha})-\log2\,.
\end{equation}
where in principle the leading order \eqref{RenyiPsi} can be plugged in. Here it is stressed that \eqref{SRREpsiEquipartition} holds exactly due to
\begin{equation}\label{EnforcedVanishing}
 \tr_R[(-1)^Fq^{nH_R}O]
 =
 0
 \quad
 \Rightarrow
 \quad
 \cZ^\phi_{(-\alpha)\alpha}(a|n)
 =
 \frac{1}{2}Z_{(-\alpha)\alpha}^\phi(q^n)
\end{equation}
for an operator $O$ containing an even number of fermions. This feature is only visible in the boundary state formalism since it allows to check all orders of entanglement entropy by drawing explicitly on the spin structure. Moreover, exact equipartition between the sectors stands in stark contrast to the bosonic case \cite{di2023boundary}, where $\Z_2$ equipartition holds only at $\cO(\epsilon^0)$.

Continuing with the remaining RDM $\rho^\psi_\aa(a)$, one decomposes the $a$-sector moments \eqref{ProjectedMoments} into spin structures according to $\tr_\aa[\Pi_aq^{H_\aa}]=\sum_{\nu=3}^4\lambda_a^{\nu}\, Z_\nu(q)$. Using the definitions of traces stated below \eqref{bdySpec} and $Z_4=\sqrt{\vartheta_4(q)/\eta(q)}=\tr_{NS}[(-1)^Fq^{H_{NS}}]$, one finds $\lambda_a^{3}=a\lambda_a^{4}=1/2$, implying
\begin{align}\label{chargeProjectedZ}
 \cZ^\psi_\aa(a|n)
 =
 \tr_\aa\left[\Pi_a(\rho^\psi_\aa)^n\right]
 =
 \frac{\sum_{\nu=3}^4\lambda_a^{\nu}\,Z_\nu(q^n)\,\Pf[\corr{\psi(w_k)\psi(w_l)}_\nu(q^n)]}{(Z_{\aa}^{\psi}(q))^n}
\end{align}
Plugging \eqref{chargeProjectedZ} into \eqref{SREE} and utilizing \eqref{PropagatorModS} leads to
\begin{align}
 S_n(\rho_\aa^\psi(\pm))
 &=
 S_n(\rho_\aa^\psi)+\frac{1}{1-n}\log\left[\frac{1\pm \X_\aa^\psi(n)}{(1\pm \X_\aa^\psi(1))^n}\right]-\log(2)\notag\\
\end{align}
where 
\begin{equation}\label{Xaa}
 \X_\aa^\psi(n)
 =
 \frac{Z_4(q^n)}{Z_3(q^n)}\frac{\Pf[\corr{\psi(w_k)\psi(w_l)}_4(q^n)]}{\Pf[\corr{\psi(w_k)\psi(w_l)}_3(q^n)]}
 \quad
\end{equation}
Recall that for $n=1$ the Pfaffian reduces to the two-point correlator. The $\log(2)$ contribution is the order of $\Z_2$ and expected on general grounds \cite{kusuki2023symmetry, Goldstein:2017bua}. When the UV cut-off shrinks, $\epsilon\to0$, then $\X_\aa(n)\simeq \tq^{\frac{1}{16n}}\to0$. Hence the leading contribution to the symmetry-resolved R\'enyi entropy stems from $S_n(\rho_\aa^\psi)$. In consequence equipartition holds to leading order, just as for entanglement in $\rho^\id_\aa$.

\subsection{Ramond Boundary States, Charged Moments and Equipartition}\label{secRoleRbkets}
The prior analysis on symmetry resolution can be phrased in a complementary way by including Ramond boundary states
\begin{align}\label{bketsR}
 \qquad
 \bket{\alpha}_R
 =
 2^{1/4}\exp\left(\iu\alpha\sum_{m=1}^\infty\psi_{-m}\bpsi_{-m}\right)\ket{\alpha}_R,
 \qquad
 (\psi_m-\iu\alpha\bpsi_{-m})\bket{\alpha}_R=0,
\end{align}
with $\alpha=\pm$. Here, $\ket{\alpha}_R$ are orthonormal Ramond sector ground states corresponding to the spin field $\sigma$ and disorder field $\mu$ of the Ising model. They have the properties
\begin{equation}\label{GonBketR}
 (-1)^F\ket{\alpha}_R=\alpha\ket{\alpha}_R\,,
 \qquad
 (-1)^{F_L}\ket{\alpha}_R=\ket{-\alpha}_R
\end{equation}
which lift to the boundary states $\bket{\alpha}_R$.  Note however that $(-1)^F$ and $(-1)^{F_L}$ are implemented projectively in the R-R sector, $(-1)^{F}(-1)^{F_L}=-(-1)^{F_L}(-1)^{F}$. In the fermionic description, $\sigma$ and $\mu$ are both viewed as fields twisted by $(-1)^F$, as reflected in the boundary state overlaps
\begin{align}
 {}_R\bbra{\alpha}\tq^{\frac{H_r}{2}}\bket{\alpha}_R
 &=
 \tr_{NS}[(-1)^Fq^{H_{NS}}]
 =
 \sqrt{\frac{\vartheta_4(q)}{ \eta(q)}}
 =
 Z_4(q)\\
 {}_R\bbra{\alpha}\tq^{\frac{H_r}{2}}\bket{-\alpha}_R
 &=
 \tr_{R}[(-1)^Fq^{H_{R}}]
 =
 0\label{RbketOverlapDistinct}
\end{align}
where $H_r$ is the bulk Hamiltonian in the R sector. In the boundary state picture, the vanishing of the second line is a simple consequence of orthogonality of the two states $\ket{\pm}_R$. Altogether, the R sector boundary states can be viewed as boundary states twisted by $(-1)^F$. For purposes of illustration, the analysis in this subsection is restricted to the invisible global state $\phi=\id$.

R sector boundary states give rise to $\Z_2$ charged moments,
\begin{align}
 Z^{\id,F}_\ab(q^n)
 =
 \tr_\ab\left[(-1)^F(\rho_\ab^\id)^n\right]
 =
 \frac{{}_R\bbra{\alpha}\tq^{\frac{1}{2n}H_r}\bket{\beta}_R}{(Z^\id_\ab(q))^n}
 =
 \frac{1}{\left(Z^\id_\ab(q)\right)^n}
 \raisebox{-.5\height}{\includegraphics[scale=.175]{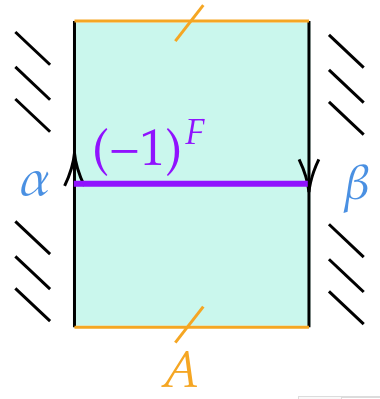}}\,.
\end{align}
Note that upper and lower edge on the figure are identified. Symmetry resolution is naturally phrased in terms of boundary states -- see for instance \cite{di2023boundary, kusuki2023symmetry} -- as seen by expressing the $a$-sector moments \eqref{ProjectedMoments} for $\phi=\id$ as follows
\begin{align}
 \cZ^\id_\ab(\pm|n)
 &=
 \frac{1}{2\left(Z^\id_\ab(q)\right)^n}\left(
 \bbra{\alpha}\tq^{\frac{H_{ns}}{2n}}\bket{\beta}\pm
 {}_R\bbra{\alpha}\tq^{\frac{H_r}{2n}}\bket{\beta}_R
 \right)\notag\\
 &=
 \frac{1}{2}\tr_\ab\left[(\rho_\ab^\id)^n\right]\left(1\pm\X_\ab^\id(n)\right)\,,\label{ProjectedMomentsBket}\\
 \X_\ab^\id(n)
 &=
 \frac{{}_R\bbra{\alpha}\tq^{\frac{H_{r}}{2n}}\bket{\beta}_R}{\bbra{\alpha}\tq^{\frac{H_{ns}}{2n}}\bket{\beta}}
\end{align}
Recall that $H_{ns}$ is the bulk Hamiltonian in the NS-NS sector and $H_r$ is the bulk Hamiltonian in the R-R sector. Due to \eqref{RbketOverlapDistinct}, $\X_{(-\alpha)\alpha}^\id=0$, leading to complete equipartion. On the other hand, when $\beta=\alpha$, the breaking of equipartion can be quantified further. Note that $\X_\aa^\id(n)$ is the analog of \eqref{Xaa}. By inserting a complete set of states in the R-R sector for the numerator and in the NS-NS sector for the denominator, the leading order in the $\tq\to0$ limit can be extracted,
\begin{align}\label{Xaa1}
 \X_\aa^\id(n)
 \simeq
 \tq^{\frac{1}{16n}}\frac{\rc_\alpha^2}{\gf_\alpha^2}
\end{align}
where $\gf_\alpha=\langle0\bket{\alpha}=1$ is the g-factor of the NS boundary state and the Ramond charge (R-charge) $\rc_\alpha={}_R\langle\alpha\bket{\alpha}_R=2^{1/4}$ is its analog in the R sector. The exponent appearing out front carries the conformal weight of the R sector ground states, $L_0\ket{\alpha}_R=\frac{1}{16}\ket{\alpha}_R$. These data quantify the breaking of equipartition as follows,
\begin{align}\label{EquipartitionBreaking}
 S_n(\rho_\aa^\id(+))-S_n(\rho_\aa^\id(-))
 &=
 \frac{1}{1-n}\log\left[
 \frac{\cZ^\id_\aa(+|n)}{\cZ^\id_\aa(-|n)}\left(\frac{\cZ^\id_\aa(-|1)}{\cZ^\id_\aa(+|1)}\right)^n
 \right]
 \notag\\
 &\overset{\tq\to0}{\simeq}
 \frac{2}{1-n}\left[\X_\aa(n)-n\X_\aa(1)\right]
 =
 \frac{2}{1-n}\frac{\rc_\alpha^2}{\gf_\alpha^2}\left(\tq^{\frac{1}{16n}}-n\,\tq^{\frac{1}{16}}\right)
\end{align}
where $\log(1+\X_\aa^\id(n))\simeq\X_\aa^\id(n)$ has been used. Therefore, the conformal dimension of the R ground state controls the order at which equipartion is broken in this system and the strength of the breaking is determined by the ratio of R-charge and g-factor. 
\section{Conformal Interfaces in the Majorana Theory}\label{secConfInt}
A different kind of global state is obtained by including non-local objects such as conformal interfaces. A large class of conformal interfaces has been constructed and analyzed for the Majorana theory in \cite{bachas2012worldsheet, bachas2013fusion} and is reviewed in this section following the discussion in \cite{Brehm:2015lja}. While \secref{secGeneralInt} begins with a description of general conformal interfaces in the Majorana theory, \secref{secBdyInt} specializes to factorizing interfaces and \secref{secTopDef} to topological defects. 

\subsection{General Majorana Theory Conformal Interfaces}\label{secGeneralInt}
Conformal interfaces $I$ are co-dimension one surfaces -- in two dimensions they simply trace out one-dimensional lines $\cC$ -- separating two possibly distinct theories, $\cft_1$ and $\cft_2$, with Hilbert space $\cH^{(i)}$, $i=1,2$. They are called conformal since they glue the energy-momentum tensors $T^{(i)},\,\bT^{(i)}$ to either side as follows
\begin{equation}\label{ConfInt}
 T^{(1)}-\bT^{(1)}
 =
 T^{(2)}-\bT^{(2)}
\end{equation}
This condition secures that one copy of the Virasoro algebra is preserved along the locus $\cC$ of the interface $I$. Folding the worldsheet along $\cC$ turns the interface into a conformal boundary condition for $\cft_1\otimes\overline{\cft_2}$, where left and right movers are interchanged, $T^{(2)}\leftrightarrow\bT^{(2)}$.

The conformal interfaces of the Majorana fermion theory constructed in \cite{bachas2012worldsheet} are parametrized by elements of $\cO\in O(2)$ and are written as operators\footnote{The same symbol is used in this text for the interface and its operator implementation.} $I(\cO):\,\cH^{(2)}\to\cH^{(1)}$ in the NS-NS or R-R sectors as follows 
\begin{align}\label{Interfaces}
 I_{NS}(\cO)=\bigotimes_{s\in\N_0+1/2}I^s(\cO)I_{NS}^0\,,
 \qquad
 I_{R}(\cO)=\bigotimes_{m\in\N_+}I^m(\cO)I_{R}^0(\cO)
\end{align}
In both sectors, the mode part takes the same shape \cite{bachas2012worldsheet},
\begin{equation}\label{Imode}
 I^s(\cO)
 =
 \exp
 \left[
 -\iu\psi_{-s}^{(1)}\cO_{11}\bpsi^{(1)}_{-s}
 +
 \psi_{-s}^{(1)}\cO_{12}\psi^{(2)}_{s}
 +
 \bpsi_{-s}^{(1)}\cO_{21}\bpsi^{(2)}_{s}
 +
 \iu\psi_{s}^{(2)}\cO_{22}\bpsi^{(2)}_{s}
 \right]
\end{equation}
Since $[I^s,I^r]=0$ for $r\neq s$, the ordering of the various $I^s$ is immaterial in \eqref{Imode}. The operator $I^m$ takes the same form, with $s\to m\in\N_+$. 

The group $O(2)$ has two disconnected components, and they can be parametrized by
\begin{align}\label{Omatrix}
 \cO_-=\begin{pmatrix}
      \cos(2\varphi) & \sin(2\varphi)\\
      \sin(2\varphi) & -\cos(2\varphi)
     \end{pmatrix}
\qquad
\cO_+=\begin{pmatrix}
      \cos(2\varphi) & \sin(2\varphi)\\
      -\sin(2\varphi) & \cos(2\varphi)
     \end{pmatrix}
\end{align}
where $\varphi\in(-\pi/2,\pi/2]$ and the subscript $\pm$ refers to their determinants, $\det\cO_\pm=\pm1$. 

Each interface has a ground state part $I(\cO)$ which in the NS-NS or R-R sector is, respectively\footnote{For convenience, labels of the Ramond ground states differ here from those in \cite{bachas2012worldsheet, bachas2013fusion, Brehm:2015lja} by a sign $\ket{\pm}_R^{\textrm{ref}}\leftrightarrow\ket{\mp}_R$. 
},
\begin{align}\label{IntZeroMode}
 I_{NS}^0=\ketbra{0}{0},
 \qquad
 I_R^0(\cO)=\sqrt{2}\left(\sin(\varphi)\ket{+}_{RR}\bra{+}+\cos(\varphi)\ket{-}_{RR}\bra{-}\right)
\end{align}
Importantly, the modes $\psi^{(1)},\,\bpsi^{(1)}$ of $\cft_1$ in \eqref{Imode} act from the left on $I_{NS,R}^0$, while the modes of $\cft_2$ act from the right. 

In the NS-NS sector Hilbert space splits according to 
\begin{equation}\label{NSbasis}
 \cH_{NSNS}=\bigotimes_{s\in\N_0+1/2}\cH_s,
 \qquad
 \cH_s=\textrm{span}\{\ket{0},\,\psi_{-s}\bpsi_{-s}\ket{0},\,\psi_{-s}\ket{0},\,\bpsi_{-s}\ket{0}\}
\end{equation}
and the operator \eqref{Imode} is most conveniently represented in this basis \cite{Brehm:2015lja},
\begin{align}\label{ImodeBasis}
 I^s(\cO)
 =
 \begin{pmatrix}
  1 & -\iu\cO_{22} & 0 & 0\\
  -\iu\cO_{11} & -\det\cO & 0 & 0\\
  0 & 0 & \cO_{12} & 0\\
  0 & 0 & 0 & \cO_{21}\\
 \end{pmatrix}_s
\end{align}
The subscript $s$ indicates that this matrix acts solely in $\cH_s$. In the R-R sector, Hilbert space splits into
\begin{equation}\label{Rbasis}
 \cH_{RR}=\bigotimes_{m\in\N_+}\cH_m^+
 \oplus
 \bigotimes_{m\in\N_+}\cH_m^-,
 \qquad
 \cH_m=\textrm{span}\{\ket{\pm}_R,\,\psi_{-m}\bpsi_{-m}\ket{\pm}_R,\,\psi_{-m}\ket{\pm}_R,\,\bpsi_{-m}\ket{\pm}_R\}
\end{equation}
 and since the modes $\psi_{-m},\,\bpsi_{-m}$ for $m\neq0$ do not intertwine the sectors built on $\ket{\pm}_R$, the operator $I^m(\cO)$ takes the same form as in \eqref{ImodeBasis}.
 
 The g-factor and R-charges of the interfaces are
 \begin{align}\label{gfactorInt}
  \gf^\cO=\bra{0}I_{NS}(\cO)\ket{0}=1\,,
  \qquad
  \rc^\cO_\alpha={}_R\bra{\alpha}I_{R}(\cO)\ket{\alpha}_R
  =\sqrt{2}
  \begin{cases}
   \sin(\varphi),\,\, \alpha=+\\
   \cos(\varphi),\,\, \alpha=-
  \end{cases}
\,.
 \end{align}

As with boundary states, the interfaces $I_{R}(\cO)$ can be viewed as the $(-1)^F$ twisted version of $I_{NS}(\cO)$. Therefore, the latter are used to construct global states below, and the former for charged moments.

Reflectivity and transmissivity are measured respectively by the coefficients
\begin{equation}\label{ReflectionTransmission}
  \cR=\cos^2(2\varphi)\,,
  \qquad
  \cT=\sin^2(2\varphi), 
  \qquad
  \cR+\cT=1\,.
\end{equation}
At $\varphi=k\pi/2$, for $k\in\Z$, the interfaces are totally reflecting, i.e. $\cR=1$, so that they are boundary conditions for $\cft_1$ and $\cft_2$. On the other hand, at $\varphi=\pm\pi/4$, the interfaces are totally transmitting $\cT=1$, and they are topological defects.
\subsection{Factorizing Interfaces}\label{secBdyInt}
Totally reflecting conformal interfaces impose proper conformal boundaries for $\cft_1$ and $\cft_2$ along the locus $\cC$, i.e. \eqref{ConfInt} becomes $T^{(1)}-\bT^{(1)} = 0 = T^{(2)}-\bT^{(2)}$. Spacetime is split into two disconnected parts and the theories factorize in the sense that they are independent (as opposed to the notion of factorization in \eqref{iota_def}). 

Factorizing interfaces are found for $\varphi=k\pi/2$, $k\in\Z$ in which case the $O(2)$ matrices \eqref{Omatrix} become diagonal, $\pm\sigma_z$ and $\pm\mathds{1}_2$, where the former is a Pauli matrix. In the basis \eqref{ImodeBasis}, these interfaces only feature an upper diagonal block, and in the mode representation \eqref{Imode} they have no mixing terms. In the NS-NS sector, it is therefore easily found that
\begin{align}\label{ObdyNS}
 I_{NS}(\Ob)
 =
 \bket{-\Ob_{11}}\bbra{\Ob_{22}}\,,
 \quad
 I_{NS}(\pm\sigma_z)
 =
 \bket{\mp}\bbra{\mp}\,,
 \quad
 I_{NS}(\pm\mathds{1}_2)
 =
 \bket{\mp}\bbra{\pm}\,.
\end{align}
This uses $\bbra{\alpha}=\bra{0}\exp\left[-\iu\alpha\sum_{s>0}\bpsi_s\psi_s\right]$. The upper sign corresponds to $\varphi=0$ and the lower sign to $\varphi=\pi/2$.

In the R-R sector, the ground state matrix \eqref{IntZeroMode} ceases to be a superposition for the factorizing values $\varphi=k\pi/2$, $k\in\Z$. One finds
\begin{align}\label{ObdyR}
 I_{R}(\pm\sigma_z)
 =
 \bket{\mp}_{RR}\bbra{\mp}\,,
 \quad
 I_{R}(\pm\mathds{1}_2)
 =
 2^{1/4}
 \bket{\mp}_{RR}\bra{\mp}\exp\left[\mp\sum_{m=1}^\infty\bpsi_m^{(2)}\psi_m^{(2)}\right]\,.
\end{align}
Parts of the $I_R(\pm\mathds{1}_2)$ interfaces do not recombine into the standard boundary states \eqref{bketsR} for $\cft_2$, because of the mismatch in signs between ${}_R\ket{\mp}$ and the exponential. This is a feature, not a bug. Below, only the interfaces based on $\pm\sigma_z$ are of interest however. As with boundary states, the interfaces $I_{R}(\pm\sigma_z)$ can be viewed as the $(-1)^F$ twisted version of $I_{NS}(\pm\sigma_z)$. 

\subsection{Topological Defects}\label{secTopDef}
Totally transmitting conformal interfaces are invisible to the energy-momentum tensor, i.e. along the locus $\cC$ \eqref{ConfInt} becomes $T^{(1)}-T^{(2)} = 0 = \bT^{(1)}-\bT^{(2)}$. 
Each component in \eqref{Omatrix} possesses two topological defects found for $\varphi=\pm\pi/4$. While $\cO_-$ contains $e$ and $(-1)^F$, $\cO_+$ contains $(-1)^{F_L}$ and $(-1)^{F_R}$,
 \begin{align}\label{Otop}
  \cO^e_-=\begin{pmatrix}
         0 & 1 \\
         1 & 0
        \end{pmatrix}\,,
 \qquad
 \cO^F_-=\begin{pmatrix}
         0 & -1 \\
         -1 & 0
        \end{pmatrix}\,,
\qquad   
\cO^{F_L}_+=\begin{pmatrix}
         0 & -1 \\
         1 & 0
        \end{pmatrix}\,,
\qquad   
\cO^{F_R}_+=\begin{pmatrix}
         0 & 1 \\
         -1 & 0
        \end{pmatrix}\,.
 \end{align}
Together they furnish the symmetry group $G=\Z_2^F\times\Z_2^{F_L}$ discussed in \secref{secSymRes}. All other conformal interfaces are marginal perturbations of these topological defects \cite{bachas2013fusion}. Observe that all topological defects are off-diagonal, $\Ot_{11}=\Ot_{22}=0$. 

In the NS-NS sector, all topological defects are implemented without obstruction, whereas in the R-R sector, the $\Ot_+$ topological defects cannot be realized by \eqref{Interfaces}. The reason is that the ground state part $I_R^0$ in \eqref{IntZeroMode} does not contain off-diagonal $\ket{\pm}_{RR}\bra{\mp}$ terms, which are necessary to represent the action of $(-1)^{F_L}$ in \eqref{GonBketR}, and similarly for $(-1)^{F_R}$. Such obstructions are expected since the group $G=\Z_2^F\times\Z_2^{F_L}$ is represented projectively in the R-R sector, which signals an 't Hooft anomaly \cite{shao2023tasi}. The R-R is the $(-1)^F$ twisted sector after all, so that defect junctions with $(-1)^{F_{L,R}}$, as drawn in \eqref{ChargedMomentI} below, are ambiguous.

Conformal interfaces require regularization when they fuse, unless one of the participating interfaces is a topological defect \cite{bachas2012worldsheet, bachas2013fusion}. Only fusion of two NS-NS interfaces is required below,
\begin{align}\label{IfuseTop}
 I_{NS}(\cO')
 &=
 I_{NS}(\cO)I_{NS}(\Ot)
 =
 \prod_{s\in\N_0+1/2}I^s(\cO')I_{NS}^0\\
 I_{NS}('\cO)
 &=
 I_{NS}(\Ot)I_{NS}(\cO)
 =
 \prod_{s\in\N_0+1/2}I^s('\cO)I_{NS}^0
\end{align}
Placement of the prime indicates the ordering of the fusion. The fused defect has a new mode operator,
\begin{align}\label{ImodeFusedTop}
 I^s(\cO')
 &=
 \begin{pmatrix}
  1 & \iu\cO_{22}\det\Ot & 0 & 0\\
  -\iu\cO_{11} & \det\cO\det\Ot & 0 & 0\\
  0 & 0 & \cO_{12}\Ot_{12} & 0\\
  0 & 0 & 0 & \cO_{21}\Ot_{21}\\
 \end{pmatrix}_s
 \\
  I^s('\cO)
 &=
 \begin{pmatrix}
  1 & -\iu\cO_{22} & 0 & 0\\
  \iu\cO_{11}\det\Ot & \,\det\cO\det\Ot & 0 & 0\\
  0 & 0 & \cO_{12}\Ot_{12} & 0\\
  0 & 0 & 0 & \cO_{21}\Ot_{21}\\
 \end{pmatrix}_s
\end{align}
By glancing at \eqref{Omatrix} and \eqref{Otop}, it is easily seen that topological defects $\Ot_+$ map between the disconnected components of $O(2)$, i.e. $I_{NS}(\cO_\pm)I_{NS}(\Ot_+)=I^s(\cO_\mp')$, while the topological defects $\Ot_-$ preserve the disconnected component,  $I_{NS}(\cO_\pm)I_{NS}(\Ot_-)=I^s(\cO_\pm')$, and similarly for inverted ordering in the fusion.

Interfaces can also be fused onto boundary states upon regularization \cite{bachas2012worldsheet}. Once more, fusion with topological defects is a regular process and serves as small consistency check with the technology employed in \secref{secSymRes}. To that end, the NS sector boundary states \eqref{bkets} are expressed in the basis \eqref{NSbasis},
\begin{equation}\label{bketNSbasis}
 \bket{\alpha}
 =
 \exp\left(\iu\alpha\sum_{s\in\N_0+1/2}\psi_{-s}\bpsi_{-s}\right)\ket{0}
 =
 \bigotimes_{s\in\N_0+1/2}\begin{pmatrix}
                           1 \\
                           \iu\alpha \\
                           0 \\
                           0 
                          \end{pmatrix}_s
\end{equation}
Fusing a topological defect onto this boundary state is easily done. Obviously, only the upper block diagonal matrix in \eqref{ImodeBasis} needs to be considered,
\begin{align}\label{IfuseTopBdy}
 I_{NS}(\Ot)\bket{\alpha}
 =
 \bigotimes_{s\in\N_0+1/2}
 \begin{pmatrix}
  1 & 0\\
  0 & -\det\Ot\\
 \end{pmatrix}_s
 \begin{pmatrix}
                           1 \\
                           \iu\alpha
                          \end{pmatrix}_s
 =
 \bket{-\det\Ot\alpha}
\end{align}
The topological defects representing $e,\,(-1)^F$ have $\det\Ot_-=-1$ and thus leave $\bket{\alpha}$ invariant, whereas the topological defects representing $(-1)^{F_{L,R}}$ have $\det\Ot_+=1$ and induce a sign flip, $\alpha\to-\alpha$, as found above in \eqref{GonBket}. In the R sector, only $\Ot_-$ are represented and act similarly, $I_{R}(\Ot_-)\bket{\alpha}_R
 =\bket{\alpha}_R$
\section{Entanglement in Global Conformal Interface States}\label{secIntEnt}
Entanglement in the presence of conformal interfaces has been investigated before in \cite{sakai2008entanglement, Brehm:2015lja, brehm2016entanglement, gutperle2016note, gutperle2017entanglement, gutperle2024non}. These works proceed by placing the replica geometry on a torus where the action of the interface on bulk degrees of freedom is naturally evaluated. In the boundary state approach to entangling edges, the system is placed on an annulus, as shown in \secref{secExcitedStates}, and therefore the analysis is qualitatively different, first and foremost due to the interplay of interfaces with boundaries.

First studies of entanglement through conformal interfaces within the boundary state approach have been conducted in \cite{karch2023universality, karch2024universal}, where universal properties such as the effective central charge have been investigated. Given the strong analytic control over interfaces in the free Majorana fermion theory, reviewed in the previous section, entanglement properties beyond R\'enyi entropies can be evaluated in full glory by applying the machinery of previous sections.

\Secref{secRDMint} constructs RDMs using conformal interfaces and presents their R\'enyi entropy. In \secref{secIntES}, the entanglement spectrum is derived for the special case that the interface is placed in the center of the interval $A$. Fermion parity resolution is applied in \secref{secSymResInt}. 

\subsection{RDMs of Global Conformal Interface States}\label{secRDMint}
\begin{figure}
\begin{center}
 \includegraphics[scale=.25]{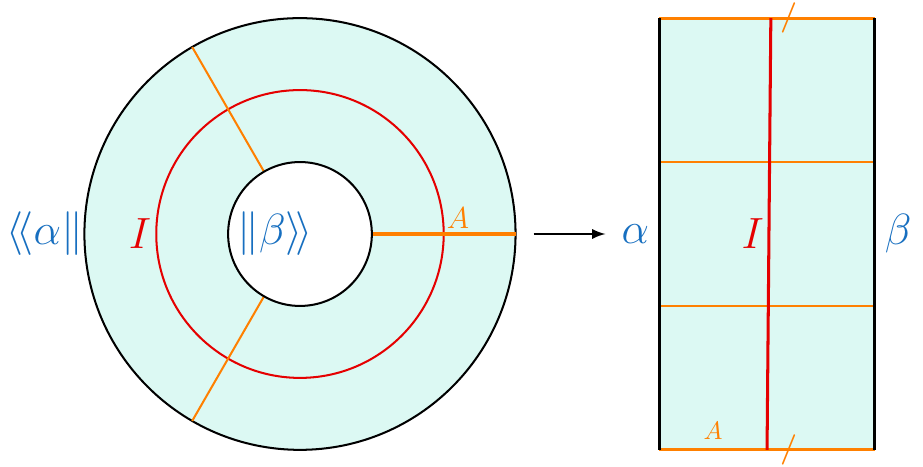}
 \end{center}
 \caption{$n=3$ replica geometry for the RDM \eqref{RDMI}. Left: on the (uniformized) complex plane. Right: on the strip.}
 \label{figEntMapInt}
\end{figure}

An RDM is constructed in analogy to \eqref{RDM}. For simplicity the interface $I$ is placed in parallel to the entangling boundaries,
\begin{equation}\label{RDMI}
  \rho_{\ab}^{I}
 =
 \frac{1}{Z^{I}_{\ab}(q)}
 \raisebox{-.5\height}{\includegraphics[scale=.15]{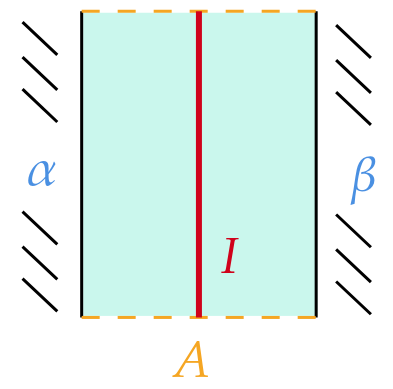}}\,,
 \qquad
 Z^{I}_{\ab}(q)
 =
 \tr_{\ab}^I\left[
 q^{H_{\ab}^I}
 \right]
 =
 \bbra{\alpha}\tq^{\frac{\delta_1}{2}H_{ns}}\,I_{NS}\,\tq^{\frac{\delta_2}{2}H_{ns}}\bket{\beta}
\end{equation}
The trace $\tr_{\ab}^I$ is over a boundary Hilbert space $\cH_\ab^I$ twisted by $I$ and $H_\ab^I$ is the Hamiltonian in this sector. To define a global state, only NS-NS sector interfaces $I_{NS}$ shall be used, so that the NS label is suppressed to avoid clutter whenever confusion cannot arise. R-R interfaces are associated naturally to the $(-1)^F$ twisted sector and become relevant when discussing charged moments, see \secref{secChargedMomentInt} below. The interface $I$ need not be placed in the center of the strip, as drawn, but is placed at a distance $\delta_1$ ($\delta_2$) from the boundary $\alpha$ ($\beta$) such that $\delta_1+\delta_2=1$. Replica geometries are constructed from \eqref{RDMI} similarly to those of local excitations $\phi$ in \figref{figEntMap}, and are depicted in \figref{figEntMapInt}

In order to evaluate the overlap $Z_\ab^I$ the propagator is expressed in the NS-NS basis \eqref{NSbasis} 
\begin{equation}\label{PropagatorAnnulus}
 \tq^{\frac{\delta}{2}H_{ns}}
 =
 \tq^{-\frac{\delta}{48}}\,
 \bigotimes_{s\in\N_0+1/2}
 \diag\left(1, \tq^{\delta s}, \tq^{\frac{\delta}{2}s}, \tq^{\frac{\delta}{2}s}\right)_s\,,
\end{equation}
which uses $H_{ns}=L_0+\bL_0-\frac{1}{24}$ as well as $[L_0,\psi_{-s}]=s\psi_{-s}$ and $[\bL_0,\psi_{-s}]=s\bpsi_{-s}$. 

In evaluating $Z^{I}_\ab$, only the first block diagonal of the mode matrices \eqref{ImodeBasis} needs to be kept. Considering an $n$-replica geometry, one obtains
\begin{align}\label{AnnulusInt}
 Z^{I}_\ab(q^n)
 &=
 \bbra{\alpha}\tq^{\frac{\delta_1}{2n}H_{ns}}\,I_{NS}(\cO)\,\tq^{\frac{\delta_2}{2n}H_{ns}}\bket{\beta}\notag\\
 &=
 \tq^{-\frac{\delta_1+\delta_2}{48n}}\bigotimes_{s\in\N_0+1/2}
 \begin{pmatrix}
  1, &-\iu\alpha
 \end{pmatrix}_s
 \begin{pmatrix}
  1 & 0 \\
  0 & \tq^{\delta_1s/n}
 \end{pmatrix}_s
 \begin{pmatrix}
  1 & -\iu\cO_{22} \\
  -\iu\cO_{11} & -\det\cO
 \end{pmatrix}_s
 \begin{pmatrix}
  1 & 0 \\
  0 & \tq^{\delta_2s/n}
 \end{pmatrix}_s
 \begin{pmatrix}
  1 \\
  \iu\beta
 \end{pmatrix}_s\notag\\
 &=
 \tq^{-\frac{1}{48n}}\prod_{s\in\N_0+1/2}
 \left[
 1
 -
 \alpha\,\cO_{11}\tq^{\delta_1s/n}
 +
 \beta\,\cO_{22}\tq^{\delta_2s/n}
 -
 \ab\det(\cO)\,\tq^{s/n}
 \right]\,.
\end{align}
Before moving on, it is easily checked that this expression reduces to \eqref{bdySpec} when employing the topological defects \eqref{Otop}, c.f. \eqref{SpinStructures}.

Because \eqref{AnnulusInt} is already in the $\tq$ channel, the R\'enyi entropy of $\rho_\ab^I$ is readily evaluated to leading order,
\begin{align}\label{RenyiInt}
 S_n(\rho^{I}_\ab)
 =
 \frac{1}{1-n}\log
 \left[
 \frac{Z^{I}_\ab(q^n)}{(Z^{I}_\ab(q))^n}
 \right]
 \overset{\tq\to0}{\simeq}
 \frac{\wi}{24}\frac{n+1}{n}+\cO(\epsilon^1)\,.
\end{align}
As in \eqref{vacuumRenyis}, the constant order vanishes. This is expected since it generally is $\log(\gf_\alpha\,\gf_\beta\,\gf^\cO)$ \cite{karch2024universal}, and $\gf_\alpha=\langle0\bket{\alpha}=\gf^\cO=1$, see \eqref{gfactorInt} and \eqref{bkets}. Therefore, in order to distinguish the RDMs $\rho_\ab^I$, and, in particular, to identify the dependence of their information content on the matrices $\cO$, their entanglement spectrum is now analyzed with care. 

\subsection{Entanglement Spectrum}\label{secIntES}
The boundary state approach allows to also evaluate the entanglement spectrum explicitly. By means of the fusion property \eqref{IfuseTop}, not all overlaps are independent,
\begin{align}\label{AnnulusIntDependencies}
 \bbra{\alpha}\tq^{\frac{\delta_1}{2}H_{ns}}\,I_{NS}(\cO_\pm)\,\tq^{\frac{\delta_2}{2}H_{ns}}\bket{-\alpha}
 &=
 \bbra{\alpha}\tq^{\frac{\delta_1}{2}H_{ns}}\,I_{NS}(\cO)I_{NS}(\Ot_+)\,\tq^{\frac{\delta_2}{2}H_{ns}}\bket{\alpha}\notag\\
 &=
 \bbra{\alpha}\tq^{\frac{\delta_1}{2}H_{ns}}\,I_{NS}(\cO_\mp')\,\tq^{\frac{\delta_2}{2}H_{ns}}\bket{\alpha}\,.
\end{align}
Hence, without loss of generality $\beta=\alpha$ is set in the following and the two distinct classes $I_{NS}(\cO_\pm)$ are investigated. That is, interest now falls on the specific overlaps
\begin{align}
 Z_\aa^{I_\pm}(q)
 =
 \bbra{\alpha}\tq^{\frac{\delta_1}{2}H_{ns}}\,I_{NS}(\cO_\pm)\,\tq^{\frac{\delta_2}{2}H_{ns}}\bket{\alpha}\,,
\end{align}
with obvious adaptation of the labelling on $Z_\aa$. In contrast to \eqref{bdySpec}, the sign $\alpha$ can now have a non-trivial effect, even though it is the same at both entangling edges, as seen from \eqref{AnnulusInt}. 

To maintain analytic control, the interface is now placed vertically in the center of the strip, $\delta_1=\delta_2=1/2$. This configuration is related to a free boson $\Z_2$ orbifold by the folding trick, as is reflected in the results below, though the folding trick is not employed directly here. Moreover, since the entanglement spectrum is sought after, the R\'enyi parameter $n$ can conveniently be set to one. Overall \eqref{AnnulusInt} reduces to 
\begin{align}
 Z_\aa^{I_\pm}(q)
 =
 \tq^{-\frac{1}{48}}\prod_{s\in\N_0+1/2}
 \left[
 1
 -
 \alpha(\,(\cO_\pm)_{11}-(\cO_\pm)_{22})\tq^{s/2}
 -
 \det(\cO_\pm)\,\tq^{s}
 \right]
\end{align}

Remarkably, for the $\cO_+$ component, the dependence on the parameter $\varphi$ of the interface disappears, as seen from \eqref{Omatrix},
\begin{equation}\label{AnnulusIntZ2}
 Z_\aa^{I_+}(q)
 =
 \tq^{-\frac{1}{48}}\prod_{s\in\N_0+1/2}
 \left[
 1
 -
 \tq^{s}
 \right]
 =
 Z_4(\tq)
 =
 Z_2(q)
 =
 Z_{\alpha(-\alpha)}^{I_-}(q)
\end{equation}
and reduces entirely to \eqref{bdySpecZ2}, pointing to the Majorana zero mode $\psi_0$. The last equality is due to \eqref{AnnulusIntDependencies}. Therefore the entanglement spectrum of the RDMs $\rho_\aa^{I_+}$ and $\rho_{\alpha(-\alpha)}^{I_-}$ is as in \eqref{ESvacuum} with ground state energy $h_i=1/16$ and has degeneracy $Z_{\psi_0}=\sqrt{2}$.

Turning to the $\cO_-$ component in \eqref{Omatrix}, the dependence on the interface modulus $\varphi$ persists,
\begin{align}
 Z_\aa^{I_-}(q)
 &=
 \tq^{-\frac{1}{48}}\prod_{s\in\N_0+1/2}
 \left[
 1
 -
 2 \alpha\,\cos(2\varphi)\tq^{s/2}
 +
 \tq^{s}
 \right]\\
 &=
 (\tq^{\frac{1}{2}})^{-\frac{1}{24}}\prod_{s\in\N_0+1/2}
 \left[
 1
 -
 \alpha\,e^{2\iu\varphi}\,\tq^{s/2}
 \right]
 \left[
 1
 -
 \alpha\,e^{-2\iu\varphi}\,\tq^{s/2}
 \right]
\end{align}
The reader is reminded that the topological defects are located at $\cos(2\varphi)=0$, i.e. $\varphi=\pm\pi/4$, for which \eqref{bdySpecZ3} is recovered as required, see \eqref{SpinStructures}. The final result depends now explicitly on the boundary conditions at the entangling edge,
\begin{subequations}\label{AnnulusIntPM}
\begin{align}
 Z_{++}^{I_-}(q)
 &=
 \frac{\vartheta_4\left(\varphi/\pi\,,\,\ttau/2\right)}{\eta\left(\frac{\ttau}{2}\right)}
 =
 q^{\frac{\varphi^2}{\pi^2}}\frac{\vartheta_2\left(-2\tau\varphi/\pi\,,\,2\tau\right)}{\eta\left(2\tau\right)}
 =
 \sum_{s\in\Z+1/2}\frac{q^{(s-\varphi/\pi)^2}}{\eta(q^2)} \\
 Z_{--}^{I_-}(q)
 &=
 \frac{\vartheta_3\left(\varphi/\pi\,,\,\ttau/2\right)}{\eta\left(\frac{\ttau}{2}\right)}
 =
 q^{\frac{\varphi^2}{\pi^2}}\frac{\vartheta_3\left(-2\tau\varphi/\pi\,,\,2\tau\right)}{\eta\left(2\tau\right)}
 =
 \sum_{m\in\Z}\frac{q^{(m-\varphi/\pi)^2}}{\eta(q^2)}
\end{align}
\end{subequations}
where the transformation properties \eqref{ChargedJacobiThetaModularS} have been employed. The factor 2 in the modular parameters results from the folding trick. Due to \eqref{AnnulusIntDependencies}, the following relations hold $Z_{++}^{I_-}(q) = Z_{+-}^{I_+}(q) = Z_{-+}^{I_+}(q)$ and $Z_{--}^{I_-}(q) = Z_{-+}^{I_+}(q) = Z_{+-}^{I_+}(q)$. Clearly, the dependence on $\alpha=\pm$ can thus be absorbed by shifting $\varphi/\pi\to\varphi/\pi+1/2$; this is not done here though.

In a variation of the results presented in \cite{cardy2016entanglement}, the entanglement Hamiltonian is given by the strip Hamiltonian $H^I_\ab$ twisted by the interface $I$,
\begin{equation}
 K_\ab^I=\frac{\pi}{\wi}H_\ab^I+\frac{1}{2\pi}\log Z_\ab^I
\end{equation}
Entanglement spectra are thus read off from \eqref{AnnulusIntPM},
\begin{subequations}\label{ESint}
\begin{align}
 \varepsilon_{++}^{I_-}(s,k)
 &=
 \frac{\pi}{\wi}\left[\left(s-\frac{\varphi}{\pi}\right)^2+2k\right]
 +
 \frac{1}{2\pi}\log Z^{I_-}_{++}\,,
 & s\in\Z+
 \frac{1}{2}\\
  \varepsilon_{--}^{I_-}(m,k)
 &=
 \frac{\pi}{\wi}\left[\left(m-\frac{\varphi}{\pi}\right)^2+2k\right]
 +
 \frac{1}{2\pi}\log Z^{I_-}_{--}\,,
 & m\in\Z\,.
\end{align}
\end{subequations}
Descendants are now counted by $2k$ with $k\in\N$ -- the factor 2 being a courtesy of the folding trick. Due to \eqref{AnnulusIntDependencies}, the following relations hold $\varepsilon_{++}^{I_-}(s,k) = \varepsilon_{+-}^{I_+}(s,k) = \varepsilon_{-+}^{I_+}(s,k)$ and $\varepsilon_{--}^{I_-}(s,k) = \varepsilon_{-+}^{I_+}(s,k) = \varepsilon_{+-}^{I_+}(s,k)$. 

The field content in \eqref{AnnulusIntPM} organizes itself into $U(1)$ representation at modular parameter $2\tau$. Because $\varphi/\pi\in(-1/2,1/2]$ the two partition functions $Z^{I_-}_{\pm\pm}$ can be rotated into each other so that they can even be regarded as representations of $\hat{\su}(2)_1$ twisted by $\varphi/\pi$, see the appendix of \cite{Gaberdiel:2011aa}. This is another instance where the folding trick shimmers through. As mentioned above, because the interface is placed in the center, the annulus overlap behaves as that of an annulus overlap without interface in the $\Z_2$ orbifold of the compact free boson. The boson annulus is half as long in the boundary state channel, i.e. instead of $\ttau$ it has $\ttau/2$, whereas the annulus is twice as long in the strip channel, i.e. instead of $\tau$ one has $2\tau$.

\subsection{Fermion Parity Resolution for Global Conformal Interface States}\label{secSymResInt}
Expressing the partition functions \eqref{AnnulusIntPM} in terms of $U(1)$ characters is in fact not natural, since the free boson orbifold has only a $\Z_2$ symmetry. More importantly, this is in accord with the symmetries of the Majorana fermion, which, as discussed in \secref{secSymRes}, is $G=\Z_2^{F}\times\Z_2^{F_L}$ and reduces to $\Z_2^F$ on the subsystem $A$. In the following, light is shed on the interplay of the conformal interface $I$ and the $\Z_2^F$ symmetry by means of fermion parity resolution. As byproduct, the spectra corresponding to the irreducible $\Z_2$ representations, in terms of which the annulus spectra should be expressed, are derived.

\Secref{secParityStab} explains how boundary states, or equivalently the factorization \eqref{iota_def}, stabilize fermion parity symmetry on conformal interfaces. \Secref{secChargedMomentInt} evaluates charged moments in presence of conformal interfaces. Fermion parity-resolved entropies are constructed in \secref{secSRREint} and their equipartition is discussed. \Secref{secFactorizingInt} concludes with a remark on factorizing interfaces.

\subsubsection{Fermion Parity Symmetry Stabilization on Conformal Interfaces}\label{secParityStab}
To begin, it needs to be checked whether the $\Z_2^F$ symmetry generated by $(-1)^F$ is a symmetry of the RDM \eqref{RDMI}, or equivalently the twisted Hamiltonian $H_\ab^I$. Similarly to the case of boundary conditions, topological endability of $(-1)^F$ on the interface needs to be confirmed. This secures that the topological defect can be moved up and down in the figure in \eqref{RDMI} without obstruction. As in the case with boundaries, topological endability of $(-1)^F$ on the interface $I$ is secured if fusion with the topological defect leaves the conformal interface invariant. Contrary to boundaries, conformal interfaces have two sides and thus left and right fusion must be considered independently. Recalling \eqref{IfuseTop} and \eqref{Otop}, it is direcly seen that
\begin{align}
 I_{NS}(\cO')
 &=
 I_{NS}(\cO)I_{NS}(\cO_-^F)
 \neq
 I_{NS}(\cO)\,,
 \\
 I_{NS}('\cO)
 &=
 I_{NS}(\cO_-^F)I_{NS}(\cO)
 \neq
 I_{NS}(\cO)
\end{align}
since the mode matrix \eqref{ImodeFusedTop} of the left- and right-fused interfaces is
\begin{align}
 I^s(\cO')
 =
 I^s('\cO)
 =
 \begin{pmatrix}
  1 & -\iu\cO_{22} & 0 & 0\\
  -\iu\cO_{11} & -\det\cO & 0 & 0\\
  0 & 0 & -\cO_{12} & 0\\
  0 & 0 & 0 & -\cO_{21}\\
 \end{pmatrix}_s
 \,
 \neq
 \,
 \begin{pmatrix}
  1 & -\iu\cO_{22} & 0 & 0\\
  -\iu\cO_{11} & -\det\cO & 0 & 0\\
  0 & 0 & \cO_{12} & 0\\
  0 & 0 & 0 & \cO_{21}\\
 \end{pmatrix}_s
 =
 I^s(\cO)
\end{align}
Because the interface $I$ is not $\Z_2^F$ symmetric, it is a priori not compatible with fermion parity resolution. One immediately notices however, that the transforming part lies purely in the lower $2\times 2$ block of $I^s$, which does not affect the boundary state $\bket{\alpha}$, see \eqref{bketNSbasis}. In fact, this is the same reason due to which $(-1)^F$ acts identically to the group unit $e$ on $\bket{\alpha}$, as shown in \eqref{IfuseTopBdy}. Hence, \textit{in acting} on $\bket{\alpha}$, the interface $I_{NS}$ is indeed $(-1)^F$-invariant,
\begin{align}
 I_{NS}(\cO)I_{NS}(\cO_-^F)\tq^{\frac{\delta}{2}H_{ns}}\bket{\alpha}
 &=
 I_{NS}(\cO)\tq^{\frac{\delta}{2}H_{ns}}\bket{\alpha}\,,
 \\
 \bbra{\alpha}\tq^{\frac{\delta}{2}H_{ns}}I_{NS}(\cO_-^F)I_{NS}(\cO)
 &=
 \bbra{\alpha}\tq^{\frac{\delta}{2}H_{ns}}I_{NS}(\cO)
\end{align}
This allows to think of the boundary as source for $(-1)^F$ topological defects and the interface as sink,
\begin{equation}\label{SourceSink}
 \raisebox{-.45\height}{\includegraphics[scale=.25]{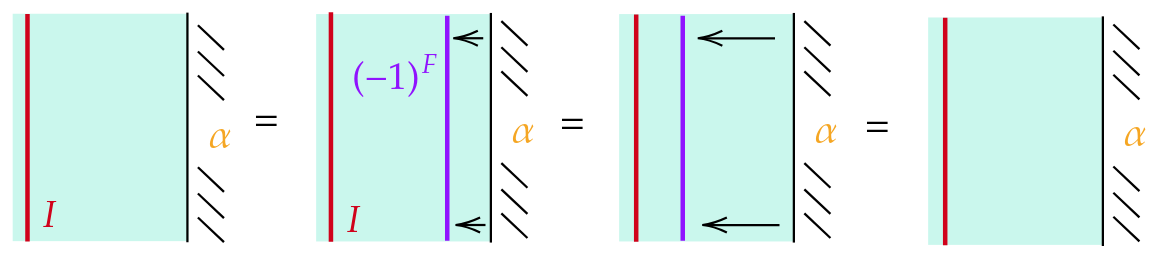}}
\end{equation}
One can say that the factorization \eqref{iota_def} \textit{stabilizes the $\Z_2^F$ symmetry against the conformal interface}. This is in stark contrast with the conventional twist field formalism, where the conformal interface simply destroys the $\Z_2^F$ symmetry otherwise present in the vacuum. 

Now that the $\Z_2^F$ symmetry is established, fermion parity resolution can be investigated. The subsystem Hilbert space splits as in \eqref{SymmetrySplitting} where $\cH_\ab^A$ is now the boundary Hilbert space $\cH_\ab^I$ twisted by $I$, and the RDM \eqref{RDMI} decomposes into block diagonal form,
\begin{align}
 \rho_\ab^I
 =
 \bigoplus_{a=\pm}\prob_a^I\rho_\ab^I(a),
 \qquad 
 \rho_\ab^I(a)
 =
 \frac{\rho_\ab^I\Pi_a}{\prob^I_a}
 =\frac{1}{\prob^I_a}
 \raisebox{-.5\height}{\includegraphics[scale=.175]{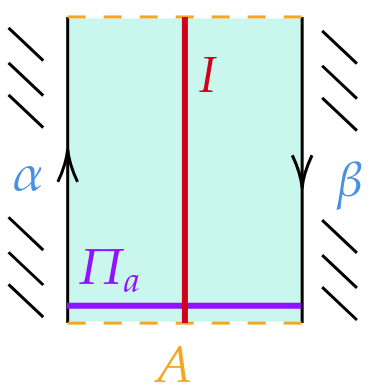}}\,,
\end{align}
via projectors $\Pi_\pm=\frac{1}{2}(e\pm(-1)^F)$ onto $V_\pm$ and probabilities $\prob_a^I=\tr_\ab^I[\Pi_a\rho_\ab^I]$. Similarly, the $a$-sector moments \eqref{ProjectedMoments} are now defined with traces $\tr_\ab^I$ over $\cH_\ab^I$. 

\subsubsection{Charged Moments}\label{secChargedMomentInt}
Because the interfaces are constructed in the bulk CFT channel, the analysis is carried out right away using boundary states as in \secref{secRoleRbkets}. To proceed, the $\Z_2^F$ charged moment
\begin{align}\label{ChargedMomentI}
 Z^{I,F}_\ab(q^n)
 =
 \tr_\ab^I\left[(-1)^F(\rho_\ab^I)^n\right]
 =
 \frac{{}_R\bbra{\alpha}\tq^{\frac{\delta_1}{2n}H_r}\,I_R(\cO)\,\tq^{\frac{\delta_2}{2n}H_r}\bket{\beta}_R}{\left(Z_\ab^I(q)\right)^n}
 =
 \frac{1}{\left(Z_\ab^I(q)\right)^n}
 \raisebox{-.5\height}{\includegraphics[scale=.16]{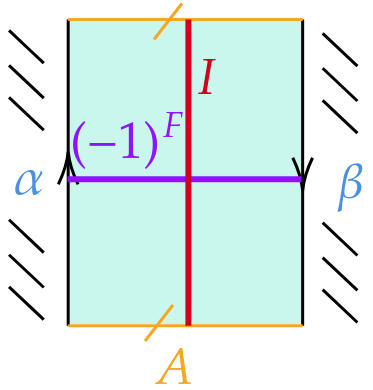}}\,.
\end{align}
is required. Note that the R sector boundary states \eqref{bketsR} and R sector interfaces in \eqref{Interfaces} appear naturally since these are the $(-1)^F$ twisted boundary states and interfaces.  In the R basis \eqref{Rbasis} the propagator takes the shape,
\begin{equation}
  \tq^{\frac{\delta}{2}H_{r}}
 =
 \tq^{\delta\left(\frac{1}{16}-\frac{1}{48}\right)}\,
 \bigotimes_{m\in\N_+}
 \diag\left(1, \tq^{\delta m}, \tq^{\frac{\delta}{2}m}, \tq^{\frac{\delta}{2}m}\right)_m
\end{equation}
so that the charged moment is straightforwardly evaluated
\begin{align}\label{ChargedMomentDeltaArbitrary}
 {}_R\bbra{\alpha}&\tq^{\frac{\delta_1}{2n}H_r}\,I_R(\cO)\,\tq^{\frac{\delta_2}{2n}H_r}\bket{\beta}_R\notag\\
 &=
 2^{1/2}\rc_\alpha^\cO\,\tq^{\frac{\delta_1+\delta_2}{24n}}\,\delta_\ab\,
 \bigotimes_{m\in\N_+}
  \begin{pmatrix}
  1, &-\iu\alpha
 \end{pmatrix}_m
 \begin{pmatrix}
  1 & 0 \\
  0 & \tq^{\delta_1m/n}
 \end{pmatrix}_m
 \begin{pmatrix}
  1 & -\iu\cO_{22} \\
  -\iu\cO_{11} & -\det\cO
 \end{pmatrix}_m
 \begin{pmatrix}
  1 & 0 \\
  0 & \tq^{\delta_2m/n}
 \end{pmatrix}_m
 \begin{pmatrix}
  1 \\
  \iu\beta
 \end{pmatrix}_m\notag\\
 &=
 2^{1/2}\rc_\alpha^\cO\,\tq^{\frac{1}{24n}}\,\delta_\ab\,\prod_{{m\in\N_+}}
 \left[
 1
 -
 \alpha(\cO_{11}\tq^{\delta_1m/n}
 -
 \cO_{22}\tq^{\delta_2m/n})
 -
 \det(\cO)\,\tq^{m/n}
 \right]
\end{align}
The $\delta_{\ab}$ arises because the interface zero mode part $I_R^0(\cO)$ in \eqref{IntZeroMode} does not contain the off-diagonal elements $\ket{\pm}_{RR}\bra{\mp}$ and furthermore ${}_R\braket{\pm}{\mp}_R=0$. The R-charge $\rc^\cO_\alpha$ of the interface is found in \eqref{gfactorInt} and $\delta_1+\delta_2=1$ has been used. Specializing once more to $\delta_1=\delta_2=1/2$, the charged moment becomes
\begin{align}
 Z^{I,F}_\ab(q^n)
 =
 2^{1/2}\rc_\alpha^\cO\,\tq^{\frac{1}{24n}}\,\delta_\ab\,\prod_{m\in\N_+}
 \left[
 1
 -
 \alpha(\cO_{11}
 -
 \cO_{22})\tq^{\frac{m}{2n}}
 -
 \det(\cO)\,\tq^{\frac{m}{n}}
 \right]
\end{align}

To proceed, the two disconnected components \eqref{Omatrix} of $O(2)$ are employed, and the notation of \secref{secIntES} is adapted for charged moments, i.e. the labelling is reduced $I_R(\cO_\pm)\to I_\pm$. Beginning with $\cO_+$ the non-vanishing charged moments are 
\begin{align}
 Z^{I_+,F}_\aa(q^n)
 &=
 2^{1/2}\rc_\alpha^\cO\,\tq^{\frac{1}{24n}}\,\prod_{m=1}^\infty
 \left[
 1
 -
 \tq^{\frac{m}{n}}
 \right]
 =
 2^{1/2}\rc_\alpha^\cO\,\eta(\tq^{1/n})
 =
 2^{1/2}\rc_\alpha^\cO\,\sqrt{-\iu n\tau}\,\eta(q^n)
\end{align}
Note that this does not reduce to \eqref{bdySpecZ2} for the topological values $\varphi=\pm\pi/4$, which is anticipated in the comments below \eqref{Omatrix}.

Turning to the $\cO_-$ component, one finds
\begin{align}
  Z^{I_-,F}_\aa(q^n)
 &=
 2^{1/2}\,\rc_\alpha^\cO\,\tq^{\frac{1}{24n}}\,\,\prod_{m=1}^\infty
 \left[
 1
 -2
 \alpha\cos(2\varphi)\tq^{\frac{m}{2n}}
 -
 \det(\cO)\,\tq^{\frac{m}{n}}
 \right]\notag\\
 &=
 2^{1/2}\,\rc_\alpha^\cO\,(\tq^{\frac{1}{2n}})^{\frac{1}{12}}\,
 \prod_{m=1}^\infty
 \left[
 1
 -
 \alpha\,e^{2\iu\varphi}\,\tq^{\frac{m}{2n}}
 \right]
 \left[
 1
 -
 \alpha\,e^{-2\iu\varphi}\,\tq^{\frac{m}{2n}}
 \right]
\end{align}
As with \eqref{AnnulusIntPM}, the result depends explicitly on $\alpha$,
\begin{subequations}\label{ChargedMomentsAA}
\begin{align}
 Z_{++}^{I_-,F}(q^n)
 &=
 \frac{\vartheta_1\left(\frac{\varphi}{\pi}\,,\,\frac{\ttau}{2n}\right)}{\eta\left(\frac{\ttau}{2n}\right)}
 =
 \iu\, q^{n\frac{\varphi^2}{\pi^2}}\frac{\vartheta_1\left(-2n\tau \frac{\varphi}{\pi}\,,\,2n\tau\right)}{\eta\left(2n\tau\right)}
 =
 \sum_{s\in\Z+\frac{1}{2}}(-1)^{s-1/2}\,\frac{q^{n(s-\varphi/\pi)^2}}{\eta(q^{2n})}\\
 Z_{--}^{I_-,F}(q^n)
 &=
 \frac{\vartheta_2\left(\frac{\varphi}{\pi}\,,\,\frac{\ttau}{2n}\right)}{\eta\left(\frac{\ttau}{2}\right)}
 =
 q^{n\frac{\varphi^2}{\pi^2}}\frac{\vartheta_4\left(-2n\tau \frac{\varphi}{\pi}\,,\,2n\tau\right)}{\eta\left(2n\tau\right)}
 =
 \sum_{m\in\Z}(-1)^{m}\,\frac{q^{n(m-\varphi/\pi)^2}}{\eta(q^{2n})}
\end{align}
\end{subequations}
where the R-charges in \eqref{gfactorInt} and \eqref{ChargedJacobiTheta} have been employed. 

\subsubsection{Fermion Parity-Resolved Entropies}\label{secSRREint}
With the charged moments at hand, the $a$-sector moments \eqref{ProjectedMoments} can be evaluated,
\begin{align}\label{ProjectedMomentsI}
 \cZ_\ab^I(\pm|n)
 =
 \tr_\ab^I\left[\Pi_\pm(\rho_\ab^I)^n\right]
 =
 \frac{1}{2(Z_\ab^I(q))^n}\left(Z_\ab^I(q^n)\pm Z_\ab^{I,F}(q^n)\right)
\end{align}

To avoid issues with projectively represented $I_R(\cO_+)$ interfaces, before symmetry resolving, global interface states based on $I_{NS}(\cO_+)$ are  first transformed into global $I_{NS}(\cO_-)$ interface states by splitting off an $I_{NS}(\Ot_+)$ interface and moving it to the boundary, as in \eqref{AnnulusIntDependencies} but read from right to left. In short $\rho_\ab^{I_+}$ is regarded as $\rho_{\alpha(-\beta)}^{I_-}$. Hence focus is reduced to global states $\rho_{\ab}^{I_-}$ in the remainder of this section.

It is easiest to start with global states $\rho_{(-\alpha)\alpha}^{I_-}$. By means of $\delta_{\ab}$ in \eqref{ChargedMomentDeltaArbitrary}, $Z_{(-\alpha)\alpha}^{I_-,F}=0$ and thus these states are completely equipartitioned,
\begin{align}
 \cZ^{I_-}_{(-\alpha)\alpha}(\pm|n)
 &=
 \frac{1}{2}\frac{Z_{(-\alpha)\alpha}^{I_-}(q^n)}{\left(Z_{(-\alpha)\alpha}^{I_-}(q)\right)^n}
 =
 \frac{1}{2}\tr_{(-\alpha)\alpha}^{I_-}\left[\left(\rho_{(-\alpha)\alpha}^{I_-}\right)^n\right]\\
 S_n\left(\rho_{(-\alpha)\alpha}^{I_-}(a)\right)
 &=
 S_n\left(\rho_{(-\alpha)\alpha}^{I_-}\right)-\log2
 =
 \frac{1}{1-n}\log\left[\frac{Z_2(q^n)}{(Z_2(q))^n}\right]-\log2
\end{align}
Recalling the relevant partition function \eqref{AnnulusIntZ2}, this comes hardly as a surprise, since it is exactly the same as \eqref{bdySpecZ2}, which is completely equipartitioned due to the Majorana zero mode. This result shows that the Majorana zero mode not only robustly enforces complete equipartition even for conformal interface states, but also hinders the interface from modifying the entanglement spectrum away from that of the vacuum. This characteristic requires only that the interface be parallel to the boundaries in the strip geometry, not that it is centered, i.e. $\delta_1\neq\delta_2$. 

Moving on to the remaining subsystem states $\rho_\aa^{I_-}$, absence of the Majorana zero mode allows the interface to have impact. The relevant partition functions \eqref{AnnulusIntPM} and charged moments \eqref{ChargedMomentsAA} are plugged into \eqref{ProjectedMomentsI}, 
\begin{subequations}\label{charactersInt}
\begin{align}
 \cZ_{++}^{I_-}(a|n)
 &=
 \frac{1}{2(Z_{++}^{I_-}(q))^n}\sum_{s\in\Z+1/2}\left(1+a(-1)^{s-1/2}\right)\,\frac{q^{n(s-\varphi/\pi)^2}}{\eta(q^{2n})}
 \equiv
 \frac{\chi^a_{++}(q^n)}{(Z_{++}^{I_-}(q))^n}\\
 \cZ_{--}^{I_-}(a|n)
 &=
 \frac{1}{2(Z_{--}^{I_-}(q))^n}\sum_{m\in\Z}\left(1+a(-1)^{m}\right)\,\frac{q^{n(m-\varphi/\pi)^2}}{\eta(q^{2n})}
 \equiv
 \frac{\chi^a_{--}(q^n)}{(Z_{++}^{I_-}(q))^n}
\end{align}
\end{subequations}
where $a=\pm1$. The character $\chi_\aa^a$ belongs to the multiplicity space $\cH_\aa^a$ of the $\Z_2$ irreducibles $V_a$ in \eqref{SymmetrySplitting} for the RDM $\rho_\aa^{I_-}$. Connecting with the comment made at the end of \secref{secIntES}, the entanglement spectrum \eqref{AnnulusIntPM} should be presented in terms of these characters instead of $U(1)$ characters. This is reflected in the resolved entropy \eqref{SREE}, which, similarly to \eqref{SRREferm}, is expressed by
\begin{align}
 S_n\left(\rho_\aa^{I_-}(a)\right)
 =
 \frac{1}{1-n}\log\left[\frac{\chi_\aa^a(q^n)}{(\chi_\aa^a(q))^n}\right]
\end{align}
Its leading orders, and in particular the breaking of equipartition, can be evaluated by returning to the $\tq$ frame in analogy to \eqref{ProjectedMomentsBket},
\begin{align}
 \cZ^{I_-}_\aa(a|n)
 &=
 \frac{1}{2\left(Z^{I_-}_\aa(q)\right)^n}
 \left(
 \bbra{\alpha}\tq^{\frac{\delta_1}{2n}H_{ns}}\,I_{NS}(\cO_-)\,\tq^{\frac{\delta_2}{2n}H_{ns}}\bket{\alpha}
 +a\,
 (NS\to R)
 \right)\notag\\
 &=
 \frac{1}{2}\tr_\aa^{I_-}\left[(\rho_\aa^{I_-})^n\right]\left(1+a\,\X_\aa^{I_-}(n)\right)\,,\\
 \X_\aa^{I_-}(n)
 &=
 \frac{{}_R\bbra{\alpha}\tq^{\frac{\delta_1}{2n}H_{r}}\,I_{R}(\cO_-)\,\tq^{\frac{\delta_2}{2n}H_{r}}\bket{\alpha}_R}{\bbra{\alpha}\tq^{\frac{\delta_1}{2n}H_{ns}}\,I_{NS}(\cO_-)\,\tq^{\frac{\delta_2}{2n}H_{ns}}\bket{\alpha}}
 \quad
 \overset{\tq\to0}{\simeq}
 \quad
 \tq^{\frac{1}{16n}}\frac{\rc_\alpha^2}{\gf_\alpha^2}\frac{\rc^\cO_\alpha}{\gf^\cO}
\end{align}
Note that the interface can be placed at any distance $\delta_{1,2}>0$ with $\delta_1+\delta_2=1$ from the boundaries in this analysis. Similarly to \eqref{Xaa1}, the ground state energy $h=1/16$ in the R-R sector, the g-factors and Ramond charges control the leading expansion of $\X_\aa^{I_-}$. Importantly, the analogs \eqref{gfactorInt} of the interface join in. 

The fermion parity resolved entropies become
\begin{align}
 S_n(\rho_\aa^{I_-}(a))
 &=
 S_n(\rho_\aa^\psi)+\frac{1}{1-n}\log\left[\frac{1+a \X_\aa^{I_-}(n)}{(1+a \X_\aa^{}(1))^n}\right]-\log(2)
\end{align}
where the entropy of the full RDM is given in \eqref{RenyiInt}. Equipartition is broken at the same order as without interface, see \eqref{EquipartitionBreaking} but with different strength,
\begin{align}\label{EquipartitionBreakingInt}
 S_n(\rho_\aa^\id(+))-S_n(\rho_\aa^\id(-))
 \overset{\tq\to0}{\simeq}
 \frac{2}{1-n}\frac{\rc_\alpha^2}{\gf_\alpha^2}\frac{\rc^\cO_\alpha}{\gf^\cO}\left(\tq^{\frac{1}{16n}}-n\,\tq^{\frac{1}{16}}\right)
\end{align}
While $\gf_\alpha=\gf^\cO=1$, $\rc_\alpha^2=\sqrt{2}$, the Ramond charge \eqref{gfactorInt} of the interface $\rc^\cO_\alpha$ can be tuned. 

\subsubsection{A Remark on Purely Reflecting Interfaces}\label{secFactorizingInt}
\begin{figure}
\begin{center}
 \includegraphics[scale=.25]{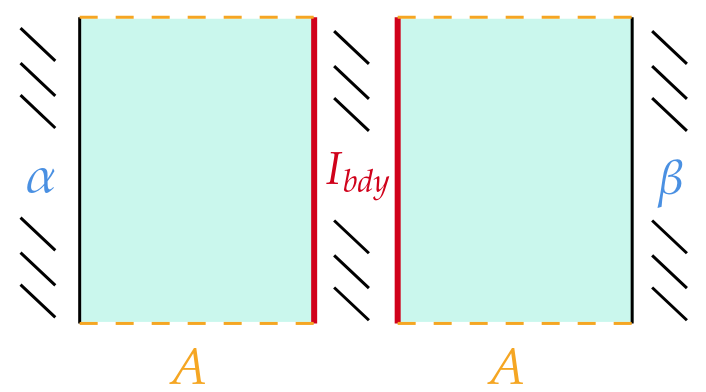}
 \end{center}
 \caption{A factorizing interface splits the entangling interval into two disconnected segments, each attached to a physical boundary resulting from the interface. The boundaries $\alpha,\beta$ describe entangling edges, as before.}
 \label{figFactorizingInt}
\end{figure}

Complete equipartition can thus be achieved for $\alpha=+1$ at $\varphi=0$ and for $\alpha=-1$ at $\varphi=\pi/2$, as is already visible in the charged moments \eqref{ChargedMomentsAA}. For these values of $\varphi$, the interface is totally reflecting \eqref{ReflectionTransmission}. Indeed, $\varphi=0$ leads to $I_{NS}(\sigma_z)$ and $\varphi=\pi/2$ leads to $I_{NS}(-\sigma_z)$ presented in \eqref{ObdyNS}. The entire setup thus describes entanglement in a product CFT each carrying a single interval attached to a physical boundary, see \figref{figFactorizingInt}.
Starting from \eqref{AnnulusInt}, the entanglement spectra describing these systems are easily confirmed to pertain to a product of BCFTs,
\begin{align}\label{productBCFT}
 Z_{++}^{I_-(\sigma_z)}(q)
 =
 Z_{--}^{I_-(-\sigma_z)}(q)
 =
 Z_4(\tq^{\delta_1})Z_4(\tq^{\delta_2})
 =
 Z_2(q^{1/\delta_1})Z_2(q^{1/\delta_2})
\end{align}
and moreover feature a Majorana zero mode. Hence
complete equipartition comes as no surprise. Due to the product structure, the two regions $A_{1,2}$ of size $\delta_{1,2}$ are independent, as is easily confirmed by consulting the mutual information $S_1(A_1)-S_1(A_1)-S_1(A_1A_2)=0$.

\section{Summary \& Outlook}\label{secOutlook}
\subsection{Summary of Results}
Entanglement in the Majorana fermion theory \eqref{action} was analyzed using the boundary state approach. Two distinct factorizations \eqref{iota_def}, determined by the spin structures \eqref{bdySpec}, were examined across three global states: the vacuum, the fermion excitation, and conformal interface states. Several information measures were employed, including the Rényi entropy, relative entropy, fermion parity-resolved entropy, and the entanglement spectrum, wherever computable.

For both the vacuum and fermion states, the Rényi entropies reduce to the corresponding results of the twist field approach, see \eqref{vacuumRenyis} and \eqref{RenyiPsi}, and are independent of the factorization induced by the spin structure. The relative entropy \eqref{RelEntropy} between the vacuum and fermion RDMs is likewise unaffected by the choice of factorization \eqref{iota_def}. Similarly, the Rényi entropies for conformal interface states reproduce the expected results anticipated in \cite{karch2024universal}. In contrast, the entanglement spectra of the subsystem vacuum states \eqref{ESvacuum} and subsystem interface states \eqref{ESint} exhibit clear dependence on both the global state and the chosen factorization. To clarify the role of this dependence, fermion parity resolution was introduced, grouping the entanglement spectrum into fermionic and bosonic sectors.

The two spin structures \eqref{bdySpec} are distinguished by the absence or presence of a Majorana zero mode, which has a decisive impact on the entanglement spectrum. When present, the zero mode enforces complete equipartition between bosonic and fermionic sectors, rather than the merely asymptotic equipartition observed otherwise. This distinction is significant: only complete equipartition signals a transition between trivial and non-trivial SPT phases and admits interpretation as a Jackiw–Rebbi domain wall. Indeed, equipartition had been observed before in the vacuum state of the Kitaev chain for large subsystems, and seen to be broken by lattice effects \cite{fraenkel2020symmetry}, thereby reflecting the CFT results obtained above. Remarkably, as shown here, this feature persists beyond the vacuum state in the fermion state and in conformal interface states. 

In the absence of a Majorana zero mode, the spectrum exhibits asymptotic equipartition, consistent with general expectations \cite{kusuki2023symmetry}. Equipartition is broken at finite order in the UV cutoff $\epsilon$. As shown in \eqref{EquipartitionBreaking} and \eqref{EquipartitionBreakingInt}, the Ramond ground-state energy determines the order at which the breaking occurs, while the Ramond charges set its magnitude. Interpreting the Ramond sector as the $(-1)^F$-twisted sector, this analysis applies more generally and can be extended to quantify equipartition breaking for other symmetry groups.

Additional formal results on the interplay of boundary states and conformal interfaces for the Majorana fermion theory were obtained along the way. Multiple partition functions for annulus overlaps including a conformal interfaces in the NS-NS and R-R sector are derived. Strikingly, symmetries which are broken by a conformal interface are shown to be stabilized in the presence of a conformal boundary state in the region delimited by both objects. Without this property, conformal interface states cannot be fermion parity-resolved. This signals a stark distinction between the boundary state approach to entanglement and the conventional twist field formalism. 

\subsection{Outlook}
Many questions for future investigation arise. Firstly, is complete equipartition generally related with non-trivial SPT phase transitions?
As explained in \cite{Cho:2016xjw}, such transitions are associated with symmetry-enforced vanishing of partition functions, which in the current framework would more specifically be symmetry-enforced vanishing of charged moments. If such vanishings are found for all but the trivial charged moment appearing in the $a$-sector moments \eqref{ProjectedMoments}, then complete equipartition is guaranteed. Once such a mechanism is explored, it remains to associate these entanglement spectra to SPT transitions. 

Secondly, does complete equipartition based on the Majorana zero mode persist in interacting fermionic minimal models? It is shown in \cite{smith2021boundary} that boundary states in fermionic minimal models fall into two classes. Whenever boundary states of distinct classes are paired, a Majorana zero mode is present in boundary spectrum. It remains to see if complete equipartition is enforced by these zero modes.

Thirdly, it is interesting to explore fermion parity resolution in supersymmetric models, since Witten indices naturally appear as charged moments, which count the difference of bosonic and fermionic ground states.

Fourthly, it is interesting to conduct more general studies of entanglement in the presence of conformal interfaces. To be able to study scenarios as in \cite{sakai2008entanglement, Brehm:2015lja, brehm2016entanglement, gutperle2016note, gutperle2017entanglement}, one needs to attach conformal interfaces to boundaries, as is done for topological defects in \cite{kojita2018topological}. However, such a procedure is expected to require regularizations. Once such a mechanism is at hand, the action of conformal interfaces on boundary fields can be accessed, allowing to evaluate information measures in their presence.

Fifthly, at present simulations in defect dressed subsystems \cite{roy2022entanglement, roy2024topological} and corresponding theoretical predictions \cite{sakai2008entanglement, Brehm:2015lja, brehm2016entanglement,  gutperle2016note, gutperle2017entanglement} are at odds. As source of discrepancy the authors of \cite{roy2022entanglement} identify the use of twisted partition functions on the torus in the theoretical predictions of \cite{sakai2008entanglement, Brehm:2015lja, brehm2016entanglement,  gutperle2016note, gutperle2017entanglement}, which is not faithful to a single interval subsystem. The framework developed here, in particular with regards to defects and interfaces, does not induce such a topology change. It is therefore promising to study if it can reproduce the behavior observed in the simulations.

Sixthly, it is worthwhile to investigate the stabilization of symmetries by factorizations in more detail. As mentioned in the main text, conformal interfaces do not respect the $\Z_2^F$ symmetry, but can nevertheless be fermion parity resolved, since the factorization \eqref{iota_def} removes the transforming components of the interface. Hence, fermion parity resolution cannot be performed in the twist field formalism. It would be interesting to test this behavior with simulations and also in other systems, for instance the free boson.

Finally, turning to holography, the standard top-down models at our disposal are supersymmetric. Hence, it is important to extend the present formalism to supersymmetry and furthermore combine this with recent advances in  holographic entanglement spectra \cite{Baiguera:2024ffx}. Holgraphic symmetry resolution \cite{Zhao:2020qmn, Weisenberger:2021eby, Zhao:2022wnp} then leads directly into studies of equipartition.  

\subsection*{Acknowledgements} I thank  Alex Belin, Aranya Bhattacharya, Jan de Boer, Horacio Casini, Shira Chapman, Saskia Demulder, Lorenz Eberhardt, Shachar Fraenkel, Giuseppe Di Giulio, Moshe Goldstein, Jessica Lopez, Pedro J. Martinez, Ingo Runkel, Amartya Singh, Herman Verlinde and in particular Paolo Rossi and Martin Schnabl for enlightening discussions. I am indebted to Giuseppe Di Giulio and Ren\'e Meyer for a careful reading of an initial version of the draft. I acknowledge hospitality at the workshop \textit{Quantum Gravity, Holography and Quantum Information} at the Federal University of Rio Grande do Norte in Natal, where part of this work was elaborated and presented. My work is funded by the European Union’s Horizon Europe Research and Innovation Programme under the Marie Skłodowska-Curie Actions COFUND, Physics for Future, grant agreement No 101081515.

\appendix

\section{Modular Forms}\label{appModularForms}
Given the modular nome $q=e^{2\pi i\tau}$, the \textit{Dedekind eta function} 
\begin{equation}\label{DedekindEta}
 \eta(q)=q^{\frac{1}{24}}\prod_{n=1}^\infty(1-q^n)
\end{equation}
appears as generating function of partitions, as character of Virasoro Verma modules.

The \textit{Jacobi theta functions} are
\begin{subequations}\label{JacobiTheta}
\begin{align}
 \vartheta_3(q)
 &=
 \sum_{n\in\Z}q^{\frac{n^2}{2}}
 =
 q^{-\frac{1}{24}}\,\eta(q)\,\prod_{n=1}^\infty\left(1+q^{n-\frac{1}{2}}\right)^2\label{JacobiTheta3}\\
 \vartheta_2(q)
 &=
 \sum_{n\in\Z}q^{\frac{1}{2}\left(n-\frac{1}{2}\right)^2}
 =
 2q^{\frac{1}{12}}\,\eta(q)\,\prod_{n=1}^\infty\left(1+q^{n}\right)^2\label{JacobiTheta2}\\
 \vartheta_4(q)
 &=
 \sum_{n\in\Z}(-1)^n\,q^{\frac{n^2}{2}}
 =
 q^{-\frac{1}{24}}\,\eta(q)\,\prod_{n=1}^\infty\left(1-q^{n-\frac{1}{2}}\right)^2\label{JacobiTheta4}\\
 \vartheta_1(q)
 &=
 i\sum_{n\in\Z}(-1)^n\,q^{\frac{1}{2}\left(n-\frac{1}{2}\right)^2}
 =
 \frac{1}{2}q^{\frac{1}{12}}\,\eta(q)\,\prod_{n=0}^\infty\left(1-q^{n}\right)^2=0\label{JacobiTheta1}
\end{align}
\end{subequations}
The second equality in all these expressions follows from the \textit{Jacobi triple product identity}
\begin{equation}
 \prod_{n=1}^\infty(1-q^n)(1+q^{n-\frac{1}{2}}w)(1+q^{n-\frac{1}{2}}w^{-1})=\sum_{m\in\Z}q^{\frac{1}{2}m^2}w^m
\end{equation}
For $\theta_3$ $w=1$ is used, for $\theta_2$ $w=q^{-1/2}$, and for $\theta_4$ $w=-1$. These product representations easily lead to 
\begin{equation}\label{tripleProduct}
 2\eta^3(q)=\theta_2(q)\theta_3(q)\theta_4(q)
\end{equation}
The following combinations are of central importance in the main text since they are naturally associated to spin structures
\begin{subequations}\label{SpinStructures}
\begin{align}
 Z_3(q)=\sqrt{\frac{\theta_3(q)}{\eta(q)}}
 &=
 q^{-\frac{1}{48}}\prod_{n=1}^\infty\left(1+q^{n-\frac{1}{2}}\right)\label{theta3eta}\\
 Z_2(q)=\sqrt{\frac{\theta_2(q)}{\eta(q)}}
 &=
 \sqrt{2}q^{\frac{1}{24}}\prod_{n=1}^\infty\left(1+q^{n}\right)\label{theta2eta}\\
 Z_4(q)=\sqrt{\frac{\theta_4(q)}{\eta(q)}}
 &=
 q^{-\frac{1}{48}}\,\prod_{n=1}^\infty\left(1-q^{n-\frac{1}{2}}\right)\label{theta4eta}\\
 Z_1(q)=\sqrt{\frac{\theta_1(q)}{\eta(q)}}
 &=
 \frac{1}{\sqrt{2}}q^{\frac{1}{24}}\,\eta(q)\,\prod_{n=0}^\infty\left(1-q^{n}\right)=0
\end{align}
\end{subequations}
The second and third line are related by modular $S$-transformation. Modular transformations act on the modular parameter as follows
\begin{equation}
 T:\quad \tau\to\tau+1,\qquad S:\quad \tau\to-\frac{1}{\tau}
\end{equation}
The modular properties of the above modular functions are
\begin{equation}\label{DedekindModular}
 \eta(\tau+1)=e^{\frac{i\pi}{12}}\eta(\tau),
 \qquad
 \eta\left(-\frac{1}{\tau}\right)=\sqrt{-i\tau}\,\eta(\tau) 
\end{equation}
and
\begin{subequations}\label{JacobiThetaModularTransf}
\begin{align}
 \vartheta_3(\tau+1)&=\vartheta_4(\tau),\qquad \vartheta_3\left(-\frac{1}{\tau}\right)=\sqrt{-i\tau}\vartheta_3(\tau)\label{JacobiTheta3Modular}\\
 \vartheta_2(\tau+1)&=e^{\frac{i\pi}{12}}\vartheta_2(\tau),\qquad \vartheta_2\left(-\frac{1}{\tau}\right)=\sqrt{-i\tau}\vartheta_4(\tau)\label{JacobiTheta2Modular}\\
 \vartheta_4(\tau+1)&=\vartheta_3(\tau),\qquad \vartheta_4\left(-\frac{1}{\tau}\right)=\sqrt{-i\tau}\vartheta_2(\tau)\label{JacobiTheta4Modular}
\end{align}
\end{subequations}
 \textit{Charged Jacobi theta functions} are defined as follows
\begin{subequations}\label{ChargedJacobiTheta}
\begin{align}
 \vartheta_3(z,\tau)
 &=
 \sum_{n\in\Z}q^{\frac{n^2}{2}}e^{2\pi \iu n z}
 =
 q^{-\frac{1}{24}}\eta(q)\prod_{r\in\N_0+1/2}(1+yq^r)(1+y^{-1}q^r)\label{ChargedJacobiTheta3}\\
  \vartheta_2(z,\tau)
  &=
  \sum_{n\in\Z}q^{\frac{1}{2}\left(n+\frac{1}{2}\right)^2}e^{2\pi \iu (n+\frac{1}{2}) z}
  =
   2\cos(\pi z)q^{\frac{1}{12}}\eta(q)\prod_{n=1}^\infty(1+yq^n)(1+y^{-1}q^n)\label{ChargedJacobiTheta2}\\
 \vartheta_4(z,\tau)
 &=
 \sum_{n\in\Z}\,q^{\frac{n^2}{2}}e^{2\pi \iu n(z+\frac{1}{2}) }
 =
  q^{-\frac{1}{24}}\eta(q)\prod_{r\in\N_0+1/2}(1-yq^r)(1-y^{-1}q^r)\label{ChargedJacobiTheta4}\\
 \vartheta_1(z,\tau)
  &=
  -\iu\sum_{n\in\Z}(-1)^nq^{\frac{1}{2}\left(n+\frac{1}{2}\right)^2}e^{2\pi \iu (n+\frac{1}{2})z}
  =
  2\sin(\pi z)q^{\frac{1}{12}}\eta(q)\prod_{n=1}^\infty(1-yq^n)(1-y^{-1}q^n)\label{ChargedJacobiTheta1}
\end{align}
\end{subequations}
where $y=e^{2\pi\iu z}$ was introduced. Their asymptotic behavior for $q\to0$ is useful, $\vartheta_3(z,\iu\infty)\to1,\,\vartheta_2(z,\iu\infty)\to2\cos(\pi z)q^{1/8},\,\vartheta_4(z,\iu\infty)\to1,\,\vartheta_1(z,\iu\infty)\to2\sin(\pi z)q^{1/8} $ The following relation is useful and easy to see,
\begin{equation}
 \p_z\vartheta_1(0,\tau)=2\pi\eta^3(q)
\end{equation}
The modular $\modS$ transformations of the charged Jacobi theta functions appear frequently and are
\begin{subequations}\label{ChargedJacobiThetaModularS}
\begin{align}
 \vartheta_3\left(\frac{z}{\tau},-\frac{1}{\tau}\right)
 &=
 \sqrt{-\iu\tau}\,e^{\iu\pi\frac{z^2}{\tau}}\vartheta_3(z,\tau)\\
  \vartheta_4\left(\frac{z}{\tau},-\frac{1}{\tau}\right)
  &=
  \sqrt{-\iu\tau}\,e^{\iu\pi\frac{z^2}{\tau}}\vartheta_2(z,\tau)\\
 \vartheta_2\left(\frac{z}{\tau},-\frac{1}{\tau}\right)
 &=
  \sqrt{-\iu\tau}\,e^{\iu\pi\frac{z^2}{\tau}}\vartheta_4(z,\tau)\\
 \vartheta_1\left(\frac{z}{\tau},-\frac{1}{\tau}\right)
  &=
  -\iu\sqrt{-\iu\tau}\,e^{\iu\pi\frac{z^2}{\tau}}\vartheta_1(z,\tau)
\end{align}
\end{subequations}

\bibliographystyle{JHEP}
\bibliography{/home/christian/Dokumente/PostdocPrag/FermionEntanglement/Bibliography.bib}

\end{document}